%% file: N2.2.nabamitaD.tex
\documentclass[11pt]{article}
\usepackage{latexsym,amssymb,amstext,amsmath,array}
\usepackage{amscd,bbm}
\usepackage{cite}
\usepackage{slashed}
\usepackage{mathrsfs}
\usepackage{hyperref}
\usepackage{xcolor}

\usepackage{xcolor}
\usepackage[english]{babel}
\usepackage[T1]{fontenc}
\usepackage[utf8]{inputenc}
\usepackage{authblk}
\usepackage{bm}
\usepackage{mathtools}
\usepackage{slashed}
\usepackage{enumerate}
\usepackage{graphicx,color}
\usepackage{cite}
\usepackage{float}
\usepackage{hyperref}
\hypersetup{
	colorlinks=true,
	linkcolor=blue,
	filecolor=magenta,      
	urlcolor=cyan,
}
\usepackage[capitalize]{cleveref}
\usepackage{verbatim}
\usepackage{lmodern}

\usepackage{setspace}

\allowdisplaybreaks
\graphicspath{{chobi/}}

\newcommand{\half}{\frac{1}{2}}

\include{sugracom}

\begin{document}

\begin{titlepage}

\begin{center}

\vskip .3in \noindent

{\Large \bf{New \emph{$\cN$=2}  SuperBMS$_3$ algebra and Invariant Dual Theory for 3D Supergravity }}

\bigskip

{Nabamita Banerjee}$^{\,a,}$\footnote{nabamita@iiserb.ac.in (on lien from IISER Pune)}, {Arindam Bhattacharjee}$^{\,b,}$\footnote{arindam.bhattacharjee@students.iiserpune.ac.in}, {Neetu}$^{\,a,} $\footnote{neetuj@iiserb.ac.in}, \\ {Turmoli Neogi}$^{\,c,}$\footnote{turmoli.neogi@ulb.ac.be}\\

\bigskip
$^{a}$ \em Indian Institute of Science Education and Research Bhopal\\
Bhopal Bypass, Bhauri, Bhopal 462066, India \\
$^{b}$ \it {Indian Institute of Science Education and
Research Pune,\\ Homi Bhabha Road, Pashan, Pune 411 008, India }\\
$^{c}$ \it{Universit\'e Libre de Bruxelles and International Solvay Institutes,\\
Campus Plain - CP 231, B-1050 Bruxelles, Belgium}\\

\vskip .5in
{\bf Abstract }
\vskip .2in
\end{center}
We have constructed a two dimensional theory dual to 3D asymptotically flat Supergravity in presence of two supercharges with(out) internal $R-$symmetry. The duals in both the cases are identified with chiral Wess-Zumino-Witten models. Further gauging the theories, we show that the dual of the one without $R-$symmetry is invariant under the well known ${\cal{N}}=2$ SuperBMS$_3$ introduced in Banerjee et.al. 1609.09210 , while for the one with $R-$symmetry, the dual is invariant under the most generic, so far unknown, quantum ${\cal{N}}=2$ SuperBMS$_3$ symmetry. We have also commented on the phase space description of the duals.    

\vfill
\eject

\end{titlepage}

\newpage
\tableofcontents
\section{Introduction and Summary}
Gravity in three space time dimensions is special, as locally it does not have any dynamical degrees of freedom. Hence, in the absence of a cosmological constant, all solutions are locally isomorphic to Minkowski spacetime $\eta_{\mu \nu}$. This feature does not make 3D gravity trivial as a large variety of gravitational solutions exists whenever global topological structures are considered. If the global topology consists of non-contractible cycles, the global solution differs from $\eta_{\mu \nu}$ (\cite{Carlip:1995zj} and references there in). It is known that 3D gravity solutions with non-trivial topology correspond to stress-energy tensors of a two dimensional theory. These two dimensional theories are usually referred as dual theory. The existence of a dual is more evident in the Chern-Simons formulation of 3D gravity \cite{WITTEN198846,Achucarro:1987vz}. 
The dual theory, in general a (chiral) Wess-Zumino-Witten model\cite{witten1984}(that we shall introduce in the next paragraph), is defined on a closed spatial section and is
obtained by solving the constraints in the Chern-Simons theory\cite{Witten:1988hf,Moore:1989yh,Elitzur:1989nr}.
In particular ordinary asymptotically flat 3D gravity can be understood as a $ISO(2,1)$ Chern-Simons gauge theory with flat boundary condition at null infinity where the Chern-Simons level $k$ is identified with Newton's constant. Here the
spatial section is a plane and the choice of boundary conditions is crucial in determining the dual theory.

It is well known that a generic Chern-Simons theory (with a compact gauge group $G$) in presence of a boundary reduces to a Wess-Zumino-Witten (WZW) model\cite{witten1984} at the boundary. The $WZW$ model is constructed by adding a non-linear sigma model (of matrix valued field $g$) in two
dimensions $\Sigma$ with a three-dimensional  $WZW$ term $\Gamma[G]$ that lives in $V$, such that $\Sigma$ is the boundary of $V$ and $G$ is the
extension of the element $g$ to $V$ \cite{Donnay:2016zka}:
\begin{equation}\label{eq1}
I_{WZW}= \frac{1}{4a^2}\int_{\Sigma}\langle \partial_{\mu}g , \partial^{\mu}(g^{-1})\rangle + \kappa \Gamma[G], \quad \Gamma[G] = \frac{1}{3} \int_V \langle G^{-1}dG, (G^{-1}dG)^2    \rangle,
\end{equation}
where $a$ and $\kappa$ are two constants.
 Although the model contains an explicit three dimensional part, its variation is two dimensional. Thus $WZW$ model describes the dynamics of two dimensional fields $g$. Such reductions have been mostly performed for asymptotically AdS 3D gravity \cite{Coussaert:1995zp,Floreanini:1987as,article,SONNENSCHEIN1988752,PhysRevLett.62.1817,Chu:1991pn,Caneschi:1996sr,Henneaux:1999ib,Valcarcel:2018kwd}. Reduction of  $ISO(2,1)$ Chern-Simons to WZW model was first studied in \cite{SALOMONSON1990769}. But we shall follow the route taken in \cite{Barnich:2013yka}, where the dual WZW model has been constructed for flat ordinary 3D gravity. In this paper other than $ISO(2,1)$ gauge algebra, the boundary conditions suitable for flat asymptotics at null infinity have been applied for the gauge field. As a result, the dual WZW model, after gauge fixing , shows invariance under infinite dimensional quantum BMS$_3$ algebra, the asymptotic symmetry of flat 3D gravity.

In this paper we shall use this construction for finding the dual of 3D asymptotically flat Supergravity theories with two supercharges. Similar analysis has been done earlier for minimal supersymmetric extension of gravity in \cite{Barnich:2015sca}. The two supercharges may rotate among themselves if an internal $R-$symmetry is present. In our study both the scenarios, absence and presence of the internal $R-$symmetry, are considered. The resultant dual for both cases corresponds to a richer chiral WZW model at the boundary. We further study the symmetries of these duals. Imposing the constraints coming from appropriate boundary conditions at null infinity, we find that the dual theory is invariant under most generic quantum ${\cal{N}}=2$ SuperBMS$_3$ symmetry. In presence of an $R$ symmetry, the ${\cal{N}}=2$ SuperBMS$_3$ algebra has three different kinds of central extensions and is so far not reported in the literature. The phase space description can be found by a Hamiltonian reduction of the models and are expected to be a generalised Liouville type theory. This will be reported in details in \cite{WIP}.

The motivation behind our construction goes as follows : the dual theory for 3D
asymptotically flat (super)gravity at null infinity is important to establish its connection with
the corresponding AdS$_3$ results\cite{HOWE1996183}. The presence of internal $R-$ charges gives a wide handle on the system. They are also crucial for the study of flat space holography in three dimensions. Most importantly these dual theories can be treated as a toy model for cosmological
scenarios \cite{Cornalba:2003kd} due to the existence of time-dependent
cosmological solutions that were found in \cite{Fuentealba:2017fck}. 

Throughout the paper, we are concerned with 3D gravity. The paper is organised as follows: in section \ref{sec1} we present the two different kinds of ${\cal{N}}=2$ SuperPoincar\'{e} algebras and their invariant bilinears. We briefly mention the 3D ${\cal{N}}=2$ Supergravity theory and its asymptotic symmetry in section \ref{sec2}. Section \ref{sec3} contains essential details about construction of a 2D dual theory of 3D flat gravity. In section \ref{sec4} we present the dual theory, i.e. ${\cal{N}}=2$ SuperPoincar\'{e} chiral WZW model. Later in sections \ref{sec5} and \ref{sec6} we study symmetries of this model. In section \ref{sec7} we present a new ${\cal{N}}=2$ SuperBMS$_3$ algebra and we conclude the paper with an outlook in section \ref{sec8}. The paper is heavy on computations and to maintain a correct flow we have presented only the important steps in the main draft. The details have been presented in six appendices that are referred at the relevant junctions in the draft.  

\section{ ${\cal{N}}=2$ SuperPoincar\'{e} algebra and Invariant Bilinears }\label{sec1}
In this paper, we are interested in finding a two dimensional theory dual to ${\cal{N}}=2$ Supergravity. As we shall see in details in later sections, to reach to our goal, we need to begin with ${\cal{N}}=2$ SuperPoincar\'{e} algebras, i.e. supersymmetric extension of Poincar\'{e} algebra with two supercharges. In this section, we shall present two distinct versions of this algebra and the invariant bilinears associated with them. These will be the building blocks of our construction.
\subsection {Two distinct ${\cal{N}}=2$ SuperPoincar\'{e} algebras} \label{sec1.1}    

There are two different versions of ${\cal{N}}=2$ SuperPoincar\'{e} algebras known in the literature \cite{HOWE1996183}. First one given as ,
\begin{eqnarray}\label{fa}
[J_{a},J_{b}]&=&\epsilon_{abc}J^{c}, \hspace{21pt} [J_{a},\mathcal{Q}_{\alpha}^{1,2}]=\frac{1}{2}(\Gamma_{a})^{\beta}_{\alpha}\mathcal{Q}_{\beta}^{1,2}, \\
\nonumber
[J_{a},P_{b}] &=&\epsilon_{abc}P^{c}, \hspace{21pt} [P_{a},\mathcal{Q}_{\alpha}^{1,2}]=0 \\
\nonumber
[P_{a},P_{b}]&=& 0, \hspace{21pt} \{\mathcal{Q}_{\alpha}^{i},\mathcal{Q}_{\beta}^{j}\}=-\frac{1}{2} (C\Gamma)^{a}_{\alpha\beta}P_{a}\delta^{ij}, 
\end{eqnarray}	
Here $J_a, P_a  {(a= 0,1,2) }$ are the Poincare generators and $\mathcal{Q}_{\alpha}^{i}$ are two distinct ${i=1,2}$ two component ${\alpha =+1,-1}$ spinors which play the role of the two fermionic generators of the algebra. The above algebra \eqref{fa} is known as ${\cal{N}}=(1,1)$ SuperPoincar\'{e} algebra. 
The other algebra is richer and it looks as ,
\begin{eqnarray}\label{N=(2,0)}
[J_a, J_b] &= \epsilon_{abc} J^c~~~~~~
[J_a, P_b] = \epsilon_{abc} P^c\\ \nonumber
[J_a, Q^i_{\alpha}] &= \frac{1}{2} (\Gamma^a)^{\beta}_{\alpha} Q^i_{\beta}~~~~~~~~
[Q^i_{\alpha},T] = \epsilon^{ij} Q^j_{\alpha}\\ \nonumber
\{Q^i_{\alpha}, Q^j_{\beta} \} &= -\frac{1}{2} \delta^{ij} (C\Gamma^a)_{\alpha \beta} P_a + C_{\alpha\beta} \epsilon^{ij} Z. 
\end{eqnarray}	
As in the previous case, $J_a, P_a$ are Poincare generators and $Q^i_{\alpha}$ are two fermionic generators and  various indices are running over same values. The important difference compared to the last case is that the two fermionic generators transform under a spinor representation of an internal R-symmetry generator $T$. As shown in \cite{HOWE1996183}, the above algebra is interesting due to the presence of a central term $Z$. This is known as ${\cal{N}}=(2,0)$ SuperPoincar\'{e} algebra. Our conventions are presented in \ref{AppA}. In this paper, we shall work with both these algebras. For the first one \eqref{fa}, our results are a trivial extension of \cite{Barnich:2014cwa}, whereas for the second one \eqref{N=(2,0)}, we get new physics , as we shall present in next sections.

\subsection {Most Generic Non-degenerate Invariant Bilinears} \label{sec1.2}
In the context of the present paper, an algebra is physically interesting when one can define a non-degenerate invariant bilinear or the quadratic Casimir for it. In the context of both the ${\cal{N}}=2$ SuperPoincar\'{e} algebras that we have written in the last section, the bilinears exist. Below we present the detailed computation for ${\cal{N}}=(2,0)$ case. \\
 For computing the bilinear, we begin with the most general quadratic combination of the generators as,
\begin{align*}
C^2 = a \eta^{ab}P_a P_b + b \eta^{ab}J_a J_b + c \eta^{ab}P_a J_b +  d_i C^{\alpha \beta} Q^i_{\alpha} Q^i_{\beta} + e  C^{\alpha \beta}\epsilon^{ij} Q^i_{\alpha} Q^j_{\beta} + f TZ + g TT + h ZZ ,
\end{align*}
where $a,b,c,d_i,e,f,g,h$ are constants that we need to determine.
For it to be a Casimir, it must commute with every generators of the algebra. An explicit computation shows that commutators of $C^2$ with $Q^i,J_c,P_c$ do not vanish while others are identically zero. Equating the four non vanishing ones to zero we get , 
$$ b=e=g=0,  \quad c = d_1 = d_2 .$$
This shows that the coefficients are fixed up to an overall factor and we fix it\footnote{It can be fixed by demanding that the bosonic Chern-Simons action reduces rightly to Einstein-Hilbert action, as we shall see in the next section.} by choosing $c=1$.
This procedure does not put any constraint on the coefficients $a$ and $h$. Thus their values can be taken to be arbitrary. Writing $C^2$ in matrix product form, we get,

\begin{eqnarray}
\nonumber
C^2= \left(\begin{matrix} P_a & J_a & Q^{1}_{\alpha} & Q^{2}_{\alpha} & T & Z\end{matrix}\right) 
 \left(\begin{matrix}
 a \eta^{ab} & \eta^{ab} & 0 & 0 & 0 & 0 \\
 \eta^{ab} & 0 & 0 & 0 & 0 & 0 \\
0 & 0 & C^{\alpha \beta} & 0 & 0 & 0 \\
 0 & 0 & 0 & C^{\alpha \beta} & 0 & 0 \\
 0 & 0 & 0 & 0 & 0 & -1\\
 0 & 0 & 0 & 0 & -1 & h\\
\end{matrix}\right)   \left(\begin{matrix} P_b \\ J_b \\ Q^{1}_{\beta}\\  Q^{2}_{\beta}\\ T \\ Z \\ \end{matrix}\right)
\end{eqnarray}

In this paper, we wish to write down Supergravity theories invariant under ${\cal{N}}=(2,0)$ and ${\cal{N}}=(1,1)$ SuperPoincar\'{e} algebra. For that purpose, we need to compute the supertrace elements between various generators. The supertrace elements come from the inverse of the above coefficient matrix. Thus, taking inverse we can write the supertrace matrix as,
\begin{eqnarray}
\nonumber
\left(\begin{matrix}
0 & \eta_{ab} & 0 & 0 & 0 & 0 \\
 \eta_{ab} & \mu \eta_{ab} & 0 & 0 & 0 & 0 \\
 0 & 0 & C_{\alpha \beta} & 0 & 0 & 0 \\
 0 & 0 & 0 & C_{\alpha \beta} & 0 & 0 \\
 0 & 0 & 0 & 0 & \bar{\mu} & -1\\
 0 & 0 & 0 & 0 & -1 & 0\\
\end{matrix}\right) ,
\end{eqnarray}
and we get the supertrace elements as, $$<J_a , P_b> = \eta_{ab}\quad <J_a, J_b> =\mu \eta_{ab}\quad <Q^I_{\alpha}, Q^J_{\beta}> = \delta^{IJ} C_{\alpha\beta}\quad <T,Z>=-1 \quad <T,T>= \bar{\mu}.$$
The arbitrariness in coefficients $a$ and $h$ manifests itself in arbitrariness of supertraces in $\langle J_a,J_b \rangle$ and $\langle T,T \rangle$ which are related by $a = \frac{1}{\mu}$, $h = \frac{1}{\,\bar{\mu}} $. One point to notice that, even for either or both of $\mu=\bar \mu=0$, the supertrace matrix is non degenerate and hence will give us a valid theory, as the one considered in \cite{Fuentealba:2017fck}\footnote{In \cite{Caroca:2018obf}, both of $\mu=\bar \mu$ were considered to be identical, but as it is clear from above analysis they are independent.} On the contrary we can not set the off diagonal elements in the first and last two blocks to zero as that will make the determinant of this matrix vanishing and hence it will be degenerate.

For the ${\cal{N}}=(1,1)$ case, we do not have the last two rows and columns and thus we get ,$$<J_a , P_b> = \eta_{ab}\quad <J_a, J_b> =\mu \eta_{ab}\quad <Q^I_{\alpha}, Q^J_{\beta}> = \delta^{IJ} C_{\alpha\beta}.$$

We shall use these supertraces in the next section.

\section{ $3$-dimensional  ${\cal{N}}=2$  Supergravity and its asymptotic symmetry}\label{sec2}

In this section, we shall study some aspects of  $3$-dimensional supergravity theories invariant under the above two symmetry algebras \eqref{fa} and \eqref{N=(2,0)}. It is well known from earlier studies \cite{Achucarro:1987vz,WITTEN198846,Barnich:2012aw} that $3$-dimensional asymptotically flat or AdS (super)gravity theories can be formulated as Chern-Simons theories. In general, Chern-Simons theory defined 
on a three dimensional manifold $M$ and invariant under the action of a compact Lie group G, is given by: 
\begin{equation}\label{csaction}
I [A] = \frac{k}{4 \pi}\int_M \langle A, dA+ \frac{2}{3}A^2 \rangle \;.
\end{equation} 
Here the gauge field $A$ is regarded as a Lie-algebra-valued one form,
and $\langle, \rangle $ represents trace using a non-degenerate invariant
bilinear form taking values on the Lie algebra space and acting as a metric and $k$ is level for the theory. Thus in a particular basis $\{T_a\}$ of the
Lie-algebra, we can express $A= A^a_{\mu}\, T_a\, {\rm d}x^{\mu}$. The equation of motion is simply given as,
\begin{equation}\label{eom} F \equiv d A + A \wedge A =0 . \end{equation}
For our purpose, we shall consider the gauge groups to be ${\cal{N}}=(1,1)$ and ${\cal{N}}=(2,0)$ SuperPoincar\'{e} groups. The 3-manifold will be a one with a {\it boundary} and we shall identify the level $k$ with Newton's constant as $ k=\frac{1}{4 G}$. For ${\cal{N}}=(1,1)$, the basis elements $\{T_a\}$ are $J_a, P_a, \mathcal{Q}_{\alpha}^{i}$, satisfying 
algebra \eqref{fa} and for ${\cal{N}}=(2,0)$, the basis elements $\{T_a\}$ are $J_a, P_a, Q_{\alpha}^{i}, T, Z$ satisfying algebra \eqref{N=(2,0)}. Using the supertrace elements as obtained in the last section we get the corresponding supergravity actions and they are respectively given as ,

\begin{equation}\label{action1}
I_{\mu,\gamma}^{(1,1)} = \frac{k}{4\pi} \int [2e^a \hat{R_a} + \mu L(\hat{\omega_a}) - \bar{\Psi}^i_{\beta} \nabla \Psi_{i}^{\beta} ] , \quad A = e^a P_a + \hat{\omega}^a J_a + \psi_i^{\alpha} \mathcal{Q}^i_{\alpha},
\end{equation}
and 
\begin{eqnarray}\label{action2}
I_{\mu,,\bar \mu,\gamma}^{(2,0)} &= \frac{k}{4\pi} \int [2e^a \hat{R_a} + \mu L(\hat{\omega_a}) - \bar{\Psi}^i_{\beta} \nabla \Psi_{i}^{\beta} -2BdC + \bar \mu BdB] ,  
\nonumber \\
&A= e^a P_a + \hat{\omega}^a J_a + \psi_i^{\alpha} Q^i_{\alpha} + BT + CZ. 
\end{eqnarray}
where $\hat{\omega}^a = \omega^a +\gamma e^a$, for some constant $\gamma$ and $\bar{\Psi}^i_{\beta} $ is the Majorana conjugate gravitino . It was first noticed in \cite{Giacomini:2006dr} that the shift in the
	spin connection is strictly needed in order to formulate this class of theories
	in terms of a Chern-Simons action. It is worth mentioning that standard ${\cal{N}}=2$ supergravity as discussed in \cite{Fuentealba:2017fck} is recovered in $\mu=\bar \mu = \gamma=0$ limit. The curvature two form $\hat R_a$, Lorentz Chern-Simons three form $L_a $ and the covariant derivative of the gravitino appearing in \eqref{action2} are respectively defined as,
\begin{align}\nonumber
\hat{R_a} =& d\hat{\omega_a} + \frac{1}{2} \epsilon_{abc}\hat{\omega^b}\hat{\omega^c}\\
L_a =& \hat{\omega^a} d\hat{\omega_a} + \frac{1}{3} \epsilon^{abc} \hat{\omega_a} \hat{\omega_b}\hat{\omega_c}\\
\nabla \Psi_{i}^{\beta} =& d \Psi_{i}^{\beta} + \frac{1}{2} \hat{\omega}^a \Psi_{i}^{\delta} (\Gamma^a)^{\beta}_{\delta} + B \Psi_{j}^{\beta} \epsilon^{ij}.\nonumber
\end{align}
Action \eqref{action1} is recovered from action  \eqref{action2} when we set the internal symmetry field parameters $B,C$ to zero. This aspect holds true for all computations and final results of the paper. Thus for the rest of the paper, to describe our results, we shall work in details  for  ${\cal{N}}=(2,0)$ group and the corresponding supergravity action \eqref{action2}.  For completion, we shall also present only results for ${\cal{N}}=(1,1)$ case in the main draft . Appendix \ref{AppC} contains computational details for this case.

\subsection{${\cal{N}}=2$ Super-BMS$_3$ Algebra}\label{sec2.1}

It is well known by now that ${\cal{N}}=2$ supergravity theories enjoy an infinite dimensional symmetry enhancement at null infinity \cite{Lodato:2016alv,Fuentealba:2017fck,Basu:2017aqn}. The asymptotic symmetry group is a 
${\cal{N}}=2$ SuperBMS$_3$ group, which is an extension of BMS$_3$ with supercharges. To get to this symmetry algebra in the Chern-Simons formulation of gravity, we need to find out a proper fall off (at null infinity) condition on the Chern-Simons gauge field. The equation of motion \eqref{eom} implies that locally the solutions of a Chern-Simons field are pure gauge $A= G^{-1}d G,$ where $G$ is a local group element. Writing the equation of motions in terms of the field parameters of \eqref{action2}, we get
\begin{align}
&d\hat{\omega} + \hat{\omega}^2 = 0, \quad
(de)^{\gamma}_{\sigma} + [\hat{\omega}, e]^{\gamma}_{\sigma} + \frac{1}{4} [\Psi_{i}^{\gamma} \bar{\Psi^{i}_{\sigma}} - \frac{1}{2} \bar{\Psi^{i}_{\beta}} \Psi_{i}^{\beta}  \delta^{\gamma}_{\sigma}] = 0\\
&d\Psi_{i}^{\beta} + (\hat{\omega} \Psi_{i})^{\beta} + B \Psi_{j}^{\beta} \epsilon^{ji} = 0, \quad
dC = \frac{1}{2} \epsilon^{ij} \Psi_{i}^{\alpha}\Psi_{j}^{\beta} C_{\alpha\beta}, \quad
dB =0
\end{align}
Where the first two equations were written after contracting the original equations with $\frac{1}{2} (\Gamma^a)$ and defining $\hat \omega= \frac{1}{2}\hat \omega^a \Gamma_a $ and $e= \frac{1}{2}e^a \Gamma_a $. The solution to these equations can be found with a bit  of algebra. The $\hat \omega$ and $B$ equations easily solve as,
\begin{align}\label{wb}
\hat{\omega} = \Lambda^{-1}d\Lambda~~~~~~~~~~
B= d\tilde{B}.
\end{align}
Coming to the spinor equations, as they are coupled, we use Jordan Decomposition method to decouple them. Defining new variables as $\mathcal{G}^1 = \frac{1}{2}(\Psi_1 -i \Psi_2)$ and $\mathcal{G}^2 = \frac{1}{2}(\Psi_1 + i \Psi_2)$ we get the new equations to be:
\begin{align}
d\mathcal{G}_1 + iB \mathcal{G}_1 + \hat{\omega} \mathcal{G}_1 = 0, \quad
d\mathcal{G}_2 - iB \mathcal{G}_2 + \hat{\omega} \mathcal{G}_2 = 0,
\end{align}
whose solutions are given as,
\begin{align}\label{g1g2}
\mathcal{G}_1 = e^{-i\tilde{B}} \Lambda^{-1} d\eta_1, \quad 
\mathcal{G}_2 = e^{i\tilde{B}} \Lambda^{-1} d\eta_2.
\end{align}
Thus the $R-$symmetry parameter field acts like a phase to the fermions. Using above results the rest of the equations of motion can be solved to give,
\begin{align}\label{ce}
C =& -i (\bar{\eta}_{1\alpha} d\eta_2^{\alpha} - \bar{\eta}_{2\alpha} d\eta_1^{\alpha} + d\tilde{C} )\\
e =& - \Lambda^{-1} [\frac{1}{2} (\eta_1 \bar{d\eta_2} + \frac{1}{2}  d\bar{\eta_2}\eta_1 \textbf{I})+ \frac{1}{2} (\eta_2 \bar{d\eta_1} + \frac{1}{2} d\bar{\eta_1} \eta_2\textbf{I}) + db] \Lambda.
\end{align}
Notice that in both of the above expressions of $C$ and $e$, the phase factors cancel among themselves.
Here $\Lambda$ is an arbitrary $SL(2,R)$ group element of unit determinant. $B, C$ are $SL(2,R)$ scalars, $\eta_i, i=1,2$ are Grassmann-valued $SL(2,R)$ spinors and $b$ is a traceless $2 \times 2$ matrix. All these are local functions of three space time coordinates $u, \phi, r$. Since we are dealing with a gauge theory, we choose a (radial) gauge condition $\partial_{\phi}A_r=0$. This implies that group element must split as $G(u,\phi,r)= g(u,\phi)h (u,r)$ and thus the gauge field must have following form, $$A=h^{-1}(a+d)h,  \qquad a= g^{-1}dg= a_u (u,\phi) du + a_{\phi} (u,\phi)d \phi.$$ We further consider that asymptotically $h= e^{-r P_0}$ and hence $\dot h(u,r)=\frac{\partial h(u,r)}{\partial u}=0$ at the boundary.
The advantage of this gauge choice is that the dependence in the radial coordinate
is completely absorbed by the group element $h$. Thus the boundary can be assumed to be unique and
located at any arbitrary fixed value of $r = r_0$, in particular to infinity. Hence, the boundary  describes a two-dimensional
timelike surface with the topology of a cylinder . Implementing the radial gauge condition, the above solutions of various field parameters can be further decomposed as\footnote{the decomposition can be obtained as, $\partial_{\phi} B_r = 0 \Rightarrow \tilde{B} = a(u,\phi) + \tilde{a}(u,r)$ and for $\partial_{\phi} \omega_r = 0 \Rightarrow \Lambda = \lambda(u,\phi) \zeta(u,r).$
	Similarly, for the fermionic fields, demanding $\partial_{\phi} \mathcal{G}^1_r =0 $ we find:\\
$	 \partial_{\phi} [e^{-i\tilde{B}} \Lambda^{-1} \partial_r \eta_1] = 0
	\Rightarrow  e^{-ia} \partial_r (\lambda^{-1} \eta_1) = \tilde{d}_1(u,r) \, (where \hspace{3pt} r-dependence \hspace{3pt} of\Lambda \hspace{3pt} is\hspace{3pt} captured \hspace{3pt}in \hspace{3pt}\tilde{d}_1)\\
	\Rightarrow   \eta_1 = e^{i a}(\lambda\tilde{d}_1 (u,r)+ d_1(u,\phi) )
$. Similarly we can find for other fields. },
\begin{eqnarray}\label{gff}
	\Lambda &=& \lambda(u,\phi) \zeta(u,r)  \nonumber \\ 
\tilde {B} &=& a(u,\phi) + \tilde{a}(u,r), \quad	\tilde{C} = c(u,\phi) + \tilde{c}(u,r) + \bar{d_2} \lambda \tilde{d_1} -  \bar{d_1} \lambda \tilde{d_2}  \nonumber \\
	\eta_1 &=& e^{i a}(\lambda\tilde{d}_1 (u,r)+ d_1(u,\phi) ), \quad 	\eta_2 = e^{-i a}(\lambda \tilde{d}_2 (u,r)+ d_2(u,\phi))\\
	b &=& \lambda E(u,r) \lambda^{-1} -\frac{1}{2} (d_1 \bar{\tilde{d_2}} \lambda^{-1} + \overline{\lambda\tilde{d_2}} d_1 \textbf{I})- \frac{1}{2} (d_2 \bar{\tilde{d_1}} \lambda^{-1} + \overline{\lambda\tilde{d_1}} d_2 \textbf{I}) + F(u,\phi), \nonumber
\end{eqnarray}
where $\dot  \zeta(u,r_0) = \dot {\tilde{a}}(u,r_0) = \dot {\tilde{c}}(u,r_0)= \dot {\tilde{d}}_1(u,r_0)= \dot {\tilde{d}}_2(u,r_0)= \dot E(u,r_0)=0.$ At the boundary, these are neither functions of $r$ nor of $u$ and must not have any dynamics. Here we see that, even onshell, the system contains arbitrary local functions $ \lambda,F,a, c,d_1,d_2$ of time $u$ (and $\phi$).  This is a common feature of a gauge theory (like for example Chern-Simons theory) that the boundary conditions and equations of motion do not uniquely fix the time ($u$) evolution of all dynamical variable. Rather a general solution of equations of motion contains arbitrary functions of time as residual degrees of freedom of the gauge system. We are looking for the theory that determines the dynamics of these residual degrees $ \lambda,F,a, c,d_1,d_2$.

Finally for ${\cal{N}}=2$ supergravity, as proposed in \cite{Fuentealba:2017fck}, the asymptotic fall of condition on the $r-$independent part of the gauge field gauge field looks like 
\begin{align}\label{bgf}
a = &\sqrt{2}[ J_1 + \frac{\pi}{k} (\mathcal{P}-\frac{4\pi}{k} \mathcal{Z}^2)J_0 + \frac{\pi}{k} (\mathcal{J} + \frac{2\pi}{k} \tau \mathcal{Z}) P_0 -\frac{\pi}{k} \psi_i Q^i_{+}  -\frac{2\pi}{k}\mathcal{Z} T - \frac{2\pi}{k} \tau Z]d\phi\\ \nonumber
& +[\sqrt{2}P_1 + \frac{8\pi}{k}\mathcal{Z}Z+ \frac{\pi}{k}(\mathcal{P}-\frac{4\pi}{k}\mathcal{Z}^2)P_0]du,
\end{align}
where various fields $\mathcal{P}, \mathcal{J},\mathcal{Z}, \tau, \psi_i$ are functions of $u,\phi$ only. These are the residual degree of freedoms and will be in correspondence to $ \lambda,F,a, c,d_1,d_2$ as introduced above in \eqref{gff}.  A technical point to note is , although 3D spacetimes can have a non trivial boundary we will not consider the holonomy terms in the following. Consequently the resulting action principle at the boundary only captures the asymptotic symmetries of
the original gravitational theory. Computing the conserved charges \cite{1974AnPhy..88..286R}, it can be shown that the asymptotic symmetry of this system is given as,
\begin{align}\label{ob2}
[M_n,J_m] &=(n-m)M_{n+m}+n^3k\;\delta_{n+m,0}, \quad 
[J_n,J_m] =(n-m)J_{n+m}\\
[M_n,R_m]&= - 4m S_{n+m}, \quad [J_n,R_m]  =-m R_{n+m} , \quad [J_n,S_m] = - m S_{n+m}\\
[R_n, S_m]&=n\;k\;\delta_{n+m,0}, \\
[J_n, \mathcal{G}^i_m]&=\;\left(\frac{n}{2}-m \right)\;\mathcal{G}^i_{n+m}, \quad (i=1,2)\\
[R_n, \mathcal{G}^1_m]&=\mathcal{G}^1_{n+m}, \quad [R_n, \mathcal{G}^2_m]=-\mathcal{G}^2_{n+m} \\
\{\mathcal{G}^{1}_n,\mathcal{G}^2_m\} &= M_{n+m}+ 2 k n^2 \delta_{n+m,0}+(n-m)S_{n+m}
\end{align}

This is the quantum symmetry algebra of \cite{Fuentealba:2017fck} presented in a diagonal basis for fermionic generators.

\section{The Boundary Theory}\label{sec3}
We are interested in constructing the two dimensional field theory that governs the dynamics of the 3D residual gauge degrees of freedom. We shall regard this as the dual theory to 3D asymptotically flat ${\cal{N}}=2$ supergravity and in this section, we shall briefly sketch this construction. Since we are interested in supergravity theories on a 3D manifold with a boundary, we need to add suitable boundary terms to the supergravity action to ensure validity of variational principle. An alternate way to look at the scenario comes from the Chern-Simons formulation of gravity. Presence of a boundary implies a non trivial fall-off conditions on the gauge fields as given in \eqref{bgf}. Hence a
 boundary term is required to add to the action in order to make solutions with the prescribed asymptotic to be a true
extrema of the action under the variational principle. For this purpose, we split the constraints coming from the boundary gauge field into two parts :
(a) constraints that relate the $u$ and $\phi$ components of the gauge field and (b) constraints that various fields of the $u$ component of the gauge field have to satisfy. 
It has been shown long back (in the context of asymptotically AdS theories) in \cite{Witten:1988hf,Moore:1989yh,Elitzur:1989nr} that pure Chern-Simons theory on a manifold with a boundary is
equivalent to a 2-dimensional chiral Wess-Zumino-Witten theory living on that boundary under conditions analogous to (a). In general, decomposing the gauge field $A(u,\phi,r)$ in time and space components as $A=duA_{u}+\tilde{A}$, the Hamiltonian form of the Chern-Simons action \eqref{csaction} can be written as\footnote{we have changed the overall sign \cite{Donnay:2016zka}} ,
\begin{eqnarray}\label{bulkaction}
I_{H}[A]&=& -\frac{k}{4\pi}\int \langle \tilde{A},\dot{\tilde{A}}du \rangle + 2\langle du A_{u},\tilde{d}\tilde{A}+\tilde{A}^{2}\rangle,
\end{eqnarray}
upto total derivatives\footnote{look at appendix \ref{AppB} for details.}. Since the fields and their derivatives do not go to zero at the boundary, for a well defined variational principle to work, we need to add $- \frac{k}{2 \pi}du \tilde d\langle A_u, \delta \tilde A \rangle$ to the Hamiltonian action. Thus the complete 2D dual theory that contains all dynamical d.o.fs of 3D gravity is governed by  
\begin{equation}
I=  I_{H}[A] - \frac{k}{2 \pi}\int du \tilde d\langle A_u, \delta \tilde A \rangle_{r=r_0}.
\end{equation}
Furthermore expressing $\tilde A= G^{-1}\tilde d G$  for some group element $G(u,r,\phi)$, the above action can be written as,
\begin{equation}\label{wzwa}
I_{WZW}=
\frac{k}{4\pi} \int_{\partial M} du d\phi \langle G^{-1}\partial_{\phi}G, G^{-1}\partial_u G \rangle -  \frac{k}{2 \pi}\int_{\partial M} du \tilde d\langle G^{-1}\partial_u G, \delta G^{-1}\tilde d G \rangle + \frac{k}{4\pi} \Gamma[G]
\end{equation}
where $\Gamma[G]$ is the three dimensional Wess-Zumino term introduced in \eqref{eq1}. 

The above action has an explicit 2D part and a 3D part $\Gamma(G)$. But the variation of this action purely lives in 2 dimensions spanned by $u, \phi$. The action \eqref{wzwa} reduces to the so called chiral Wess-Zumino-Witten model that is dual to a 3D Chern-Simons theory with a boundary. In the subsequent sections, we shall construct such a  Wess-Zumino-Witten model and study its symmetry properties. As we shall see, after incorporating the radial gauge fixing conditions, the dynamics will only depend on two dimensional fields.

\section{${\cal{N}}=2$ SuperPoincar\'{e} Wess-Zumino-Witten model}\label{sec4}
In this section, we shall write down the two dimensional Wess-Zumino-Witten (WZW) model dual to ${\cal{N}}=(2,0)$ supergravity defined at null infinity $r \rightarrow \infty$ hypersurface of the bulk 3D spacetime. 
Following the prescription outlined in the last section, we first write down (a) type of constraints on the asymptotic gauge field \eqref{bgf}, relating its $u$ and $\phi$ components as , 
\begin{align}\label{bc1}
e^a_u = \omega^a_{\phi}, ~~~~~~~\omega^a_u =0, ~~~~~~~~~\psi^{\pm}_{Iu}=0, \qquad
B_u = 0,~~~~~~-4B_{\phi} = C_u.
\end{align}
The $u$ component of the gauge field \eqref{bgf} is further constrained and we shall come back to this point later.
Under these constraints \eqref{bc1} the surface term at the boundary looks like:\\
\begin{equation}
I_{surf} =  - \frac{k}{2 \pi}\int du \tilde d\langle A_u, \delta \tilde A \rangle_{r_0 \rightarrow \infty}=-\frac{k}{4\pi} \int_{\partial M} du d\phi [\omega^a_\phi \omega_{a\phi} + 4 B^{2}_{\phi}] _{r_0 \rightarrow \infty},
\end{equation}
where the $\phi-$ total derivative has been set to zero.
Using the field parameters as defined in \eqref{action2} and the supertrace elements the total action \eqref{wzwa} can be expressed as:
\begin{align}\label{sa}
I_{(2,0)}=&\frac{k}{4\pi} \bigg[\int du d\phi [e^a_{\phi} \omega_{au} + \omega^a_{\phi} e_{au} + \mu \omega^a_{\phi}\omega_{au} + \bar{\psi}^u_{i\alpha} \psi^{\alpha}_{i\phi} - B_{\phi} C_u - C_{\phi} B_u - \omega^a_{\phi} \omega_{a\phi} - 4 B^{\phi} B_{\phi}+ \bar{\mu} B_{\phi} B_u]_{r_0 \rightarrow \infty}\nonumber \\
&+\frac{1}{6} \int [3 \epsilon_{abc} e^a \omega^b \omega^c + \mu \epsilon_{abc} \omega^a \omega^b \omega^c  + \frac{3}{2} \omega^a (C\Gamma_a)_{\alpha\beta} \Psi_i^{\alpha} \psi_i^{\beta} + 3B \psi_i^{\alpha} \psi_j^{\beta} C_{\alpha\beta} \epsilon^{ij}]\bigg]
\end{align}
As has been discussed in section \ref{sec2}, in an onshell gauge systems, there are left over residual degrees of freedom. To get  the theory (action) that describes the dynamics of these degrees of freedom, we first evaluate the above action on the solutions of equations of motions obtained in section \ref{sec2} as:
\begin{align}\label{wzwa2}
I_{(2,0)}=&\frac{k}{4\pi} \bigg[\int du d\phi Tr[2\mu \Lambda^{-1} \Lambda' \Lambda^{-1} \dot{\Lambda} -2( {\bar{\eta_1}}' \dot{\eta_2}+{\bar{\eta_2}}' \dot{\eta_1}) + 2i \dot{\tilde{B}} (\bar{\eta}_{1} \eta_2 - \bar{\eta}_{2} \eta_1) - 2 (\Lambda^{-1} \Lambda')^2 - 4 (\tilde{B}')^2  \nonumber \\
& -4 \dot{\Lambda} \Lambda^{-1} (\frac{1}{2} (\eta_1 \bar{\eta}'_2 + \bar{\eta_2} \eta'_1 \textbf{I}) + \frac{1}{2} (\eta_2 \bar{\eta}'_1 + \bar{\eta_1}\eta'_2  \textbf{I}) + b')+ 2i \dot{\tilde{B}} \tilde{C}' + \bar{\mu} \tilde{B}' \tilde{\dot{B}}]+ \frac{2\mu}{3} \int Tr[(d\Lambda \Lambda^{-1})^3]\bigg]
\end{align}
Let us briefly mention the origin of various terms appearing in \eqref{wzwa2}. The terms in \eqref{sa} proportional to $\mu, \bar\mu$ directly reduces to their counterpart in \eqref{wzwa2} onshell whereas the term $3B \psi_i^{\alpha} \psi_j^{\beta} C_{\alpha\beta} \epsilon^{ij}$ gives rise to a 2D piece which added with the three other boundary pieces $- B_{\phi} C_u - C_{\phi} B_u - 4 B^{\phi} B_{\phi}$ gives the terms $2i \dot{\tilde{B}} (\bar{\eta}_{1\alpha} \eta_2^{\alpha'}- \bar{\eta}_{2\alpha} \eta_1^{\alpha'})- 4 (\tilde{B}')^2 +2i \dot{\tilde{B}} \tilde{C}'$.
In a similar way, the bulk terms $3 \epsilon_{abc} e^a \omega^b \omega^c $ onshell gives a 2D piece which clubbed with boundary terms $e^a_{\phi} \omega_{au} + \omega^a_{\phi} e_{au}$ gives $-4 \dot{\Lambda} \Lambda^{-1} (\frac{1}{2} (\eta_1 \bar{\eta}'_2 + \bar{\eta_2} \eta'_1 \textbf{I}) + \frac{1}{2} (\eta_2 \bar{\eta}'_1 + \bar{\eta_1}\eta'_2  \textbf{I}) + b')$. Finally $\frac{3}{2} \omega^a (C\Gamma_a)_{\alpha\beta} \Psi_i^{\alpha} \psi_i^{\beta}$ just vanishes onshell up to total derivatives. The terms proportional to $I$ in \eqref{wzwa2} actually give zero contributions.

Further using the gauge decomposed forms of the solutions as in \eqref{gff} and neglecting total derivatives in $u,\phi$, the above action rightly simplifies to,
\begin{align}\label{gfa}
I_{(2,0)}=&\frac{k}{4\pi}\bigg\{ \int du d\phi Tr[2\mu \lambda^{-1} \lambda' \lambda^{-1} \dot{\lambda} -2( {\bar{d_1}}' \dot{d_2}+{\bar{d_2}}' \dot{d_1}) - 2i {a}' (\bar{d}_{1} \dot{d_2} - \bar{d}_{2}\dot{d_1} + \dot{\lambda} \lambda^{-1}(d_2 \bar{d_1} - d_1 \bar{d_2}) ) \nonumber\\ &- 4 (a')^2 
 -2 \dot{\lambda} \lambda^{-1} ( d_1 \bar{d}'_2  + d_2 \bar{d}'_1 +2 F') - 2 (\lambda^{-1} \lambda')^2+ 2i \dot{a} c' + \bar{\mu} a' \dot{a}]\nonumber\\&+ \frac{2\mu}{3} \int Tr[(d\Lambda \Lambda^{-1})^3]\bigg\},
\end{align}
where we have dropped terms proportional to $I$ in \eqref{gfa} as they actually give zero contributions.
It is interesting to note that the dependence on $\zeta ,\tilde{a},\tilde{c},\tilde{d}_1,\tilde{d}_2, E$ drops from the  two dimensional part of the last expression. Also, 
as physically expected, \eqref{gfa} is exactly same as \eqref{wzwa2} when we replace various onshell fields $\Lambda,b,\tilde{B},\tilde{C},\eta_1,\eta_2$ by their non trivial gauge fixed counterpart $ \lambda,F,a, c,d_1,d_2$ respectively as given in \eqref{gff}. One can easily check that the variation of action \eqref{gfa} is purely two dimensional
\footnote{In particular the variation of the last 3D term is given as 
$$ \frac{1}{3}\frac{\delta \int Tr[(d\Lambda \Lambda^{-1})^3]}{\delta \lambda^{\alpha}_{\beta} }= [(\lambda^{-1})' \dot \lambda \lambda^{-1} - (\lambda^{-1})^{{\mathbf{\cdot}}} \lambda ' \lambda^{-1} ]^{\beta}_{\alpha}$$}. This is the chiral Wess-Zumino-Witten (WZW) model dual to 3D asymptotically flat ${\cal{N}}=(2,0)$ Supergravity and is the main result of the paper. Similarly the chiral WZW model dual to  3D asymptotically flat ${\cal{N}}=(1,1)$ Supergravity takes the following form,
\begin{align}\label{gfa1}
I_{(1,1)} &=\frac{k}{2\pi}\bigg\{ \int du d\phi[2\dot{\lambda}\lambda^{-1}\alpha'+\half\sum_{i=1}^{2}\dot{\lambda}\lambda^{-1}\nu^{i}\overline{\nu}^{i '}-(\lambda' \lambda^{-1})^{2}+\mu\lambda' \lambda^{-1}\dot{\lambda}\lambda^{-1}-\half\sum_{i=1}^{2}\dot{\overline{\nu}}^{i}\nu^{i '}] \nonumber \\
&+\frac{\mu}{3}\int Tr(d\Lambda\Lambda^{-1})^{3}\bigg\},
\end{align}
where various fields are defined in the appendix \ref{AppC}. 

To further analyse the dynamics of the above two dimensional theory \eqref{gfa}, let us first write down the equations of motion of various fields. They are given as,
\begin{align}\label{geom}
eom~of F:\quad \qquad &(\dot{\lambda} \lambda^{-1})'=0\\
eom~of c:\quad \qquad &(\dot{a})'=0\\
eom~of d_1:\quad \qquad &\dot{\bar{d'_2}} + \bar{d'_2} (\dot{\lambda} \lambda^{-1}) - i a'(\dot{ \bar{d_2}}+ \bar{d_2} \dot{\lambda} \lambda^{-1}) =0\\
eom~of d_2:\quad \qquad  &\dot{\bar{d'_1}} + \bar{d'_1} (\dot{\lambda} \lambda^{-1}) + i a'(\dot{ \bar{d_1}}+ \bar{d_1} \dot{\lambda} \lambda^{-1}) = 0\\
eom~of a:\quad \qquad &\dot{c'} - ( {\bar{d_1}} \dot{d_2} -  {\bar{d_2}} \dot{d_1} +\dot{\lambda} \lambda^{-1}(d_2 \bar{d_1} - d_1 \bar{d_2}) )' + 4i a'' = 0\\
eom~of \lambda: \quad \qquad  &\dot{\hat\alpha} - (\dot{\lambda} \lambda^{-1}) \hat\alpha + \hat\alpha (\dot{\lambda} \lambda^{-1}) + 2(\lambda' \lambda^{-1})' =0
\end{align}

where we have defined $\hat\alpha = 2F' + d_2 (\bar{d_1}' + ia' \bar{d_1}) + d_1 (\bar{d_2}' - ia' \bar{d_2})$. The above equations are simplified versions of original equations of motion, where for simplifying one equation we have iteratively used other ones. For example, in the last equation the $\mu$ term drops off with a careful calculation and use of the first equation\footnote{There are two typos in equation (24) and (25) of \cite{Barnich:2015sca} related to the $\mu$ dependent term. The contribution to the eom from $\mu$ term is 
	$~  (\lambda^{-1} \lambda ')^{{\mathbf{\cdot}}}= \lambda^{-1} (\dot \lambda \lambda^{-1})' \lambda^{-1}=0$ by using first equation.}. 
Similarly,we have used the  $c$ eom in deriving the final eom of $a$. This eliminates the $\bar{\mu}$ dependent piece .
Thus we see that the final forms of eom do not contain any of the unfixed supertrace elements $\mu$ or $\bar{\mu}$. Hence the solutions of these fields will also be independent of these parameters. 
The generic solutions of these equations further decompose into individual functions of $u$ and $\phi$ and are given as,
\begin{align}\label{gsol}
\lambda &= \tau(u) \kappa(\phi)\\
a &= a_1(u) + a_2(\phi)\\
d_1&= e^{-ia_2}\tau (\zeta_1^{(1)} (u)+ \zeta_2^{(1)} (\phi) )\\
d_2 &= e^{ia_2}\tau (\zeta_1^{(2)} (u)+ \zeta_2^{(2)} (\phi) )\\
c &=   \bar{\zeta_2}^{(1)} \zeta_1^{(2)} - \bar{\zeta_2}^{(2)} \zeta_1^{(1)} + c_1(u) + c_2(\phi) - 4iua_2'\\
F &= \tau [a_F(\phi) + \delta_F(u) - u \kappa' \kappa^{-1} -\frac{1}{2} ( \zeta_1^{(2)} \bar{\zeta_2}^{(1)}+ \zeta_1^{(1)} \bar{\zeta_2}^{(2)})] \tau^{-1}
\end{align}

As it turns out, this chiral WZW model is invariant under rich symmetries. In the next subsection, we shall study these symmetries and their consequences.

\section{Symmetries of The Chiral WZW Model}\label{sec5}
\subsection{Global Symmetry}\label{sec5.1}
The Chiral WZW model \eqref{wzwa2} that we derived in the last section in invariant under a set of global symmetries. Various fields enjoy a coordinate $u,\phi$ dependent transformation under these symmetries and they are given as,  
\begin{align*} 
a &\rightarrow a + A(\phi); \quad c\rightarrow c- 4iu A'; \quad
d_1 \rightarrow e^{-iA} d_1; \quad d_2 \rightarrow e^{iA} d_2\\
c &\rightarrow c + \mathcal{C}(\phi)\\
\lambda &\rightarrow \lambda \theta^{-1}(\phi); \quad F \rightarrow F + u \lambda ( \theta^{-1}\theta') \lambda^{-1}\\
F &\rightarrow F + \lambda N(\phi) \lambda^{-1}\\
d_1 &\rightarrow d_1 + \lambda D_1(\phi); \quad c \rightarrow c + \bar{D_1}(\phi) \lambda^{-1} d_2; \quad F \rightarrow F -\frac{1}{2} d_2 \bar{D_1}(\phi) \lambda^{-1}\\
d_2 &\rightarrow d_2 + \lambda D_2(\phi); \quad c \rightarrow c - \bar{D_2}(\phi) \lambda^{-1} d_1; \quad F \rightarrow F - \frac{1}{2} d_1 \bar{D_2}(\phi) \lambda^{-1}
\end{align*}
In each of the above expressions, the fields that are not written remain unchanged under that corresponding symmetry transformation. Thus, we find that there are six finite symmetry transformations, generated by scalar parameters $A(\phi),\mathcal{C}(\phi)$, matrix valued parameters $\theta(\phi),N(\phi)$ and spinor parameters $D_1(\phi),D_2(\phi)$.  

One possible way to get to these symmetry transformations is to look for the symmetries of the solutions given in \eqref{gsol}\footnote{We are thankful to Prof. Glenn Barnich for clarifying this point to us.}. We have presented relevant details of the computations in appendix \ref{AppD}. Once we obtain these transformations, they can be proved to be symmetries of the action as well. 

Next we look for the Noether currents corresponding to the above symmetries. For this purpose, we need the infinitesimal versions of the above symmetry transformations that are as follows,
\begin{align} \label{ist}
\delta_A a &= A(\phi); \quad \delta_A c = - 4iu A'; \quad
\delta_A d_1 = -iA d_1; \quad \delta_A d_2 = iA d_2 \\
\delta_\mathcal{C}c &= \mathcal{C}(\phi)\\
\delta_\theta\lambda &= -\lambda \Theta(\phi); \quad \delta_\theta F =+ u \lambda \Theta'  \lambda^{-1}\\
\delta_N F &= \lambda N(\phi) \lambda^{-1}\\
\delta_{D1}d_1 &= \lambda D_1(\phi); \quad \delta_{D1}c = \bar{D_1}(\phi) \lambda^{-1} d_2; \quad \delta_{D1} F = -\frac{1}{2} d_2 \bar{D_1}(\phi) \lambda^{-1}\\
\delta_{D2}d_2 &= \lambda D_2(\phi); \quad \delta_{D2}c= -\bar{D_2}(\phi) \lambda^{-1} d_1; \quad \delta_{D2} F = -\frac{1}{2} d_1 \bar{D_2}(\phi) \lambda^{-1}.
\end{align}
Here we have used same $A(\phi),\mathcal{C}(\phi),N(\phi),D_1(\phi),D_2(\phi)$ as infinitesimal transformation parameters and $\Theta (\phi)$ is the infinitesimal parameter for $\theta$ transformation as $\theta(\phi)= I + \Theta (\phi)$.
For a theory $S[\phi_i]= \int d^x {\cal{L}}(\phi_i, \partial_{\mu}\phi_i)$, the Noether current associated to a global symmetry generated by parameter $\epsilon$ is given as, $${\cal {J^{\mu}}}_{\epsilon}= \frac{\delta \cal{L}}{\delta(\partial_{\mu} \phi_i)}\delta_{\epsilon}\phi_i- K ^{\mu}_{\epsilon}, \qquad \partial_{\mu}K ^{\mu}_{\epsilon}= \delta_{\epsilon}\cal{L}.$$
The current is conserved i.e. $\partial_{\mu}{\cal {J^{\mu}}}_{\epsilon}=0$ onshell. From this definition there is a clear ambiguity in the identification of current, as  \begin{equation}\label{nci}
J^{\mu}_{\epsilon} \sim {\cal {J^{\mu}}}_{\epsilon} + \partial_{\nu}S^{[\mu\nu]}_{\epsilon}+T^{\mu}_{\epsilon},
\end{equation}
where $S^{[\mu\nu]}_{\epsilon}$ is any antisymmetric tensor in its indices and $T^{\mu}_{\epsilon}$ is a possible vector that is onshell divergenceless. Both the currents will generate the same symmetry for the system.
Below we present the currents corresponding to above symmetries \eqref{ist},
\begin{align}\label{cc}
J^{\mu}_{A} &=\delta^{\mu}_{0} \frac{k}{4\pi}Tr [2 \bar{\mu} a' + 2i c' - 8u a''+2i(\bar{d_2}' d_1 - \bar{d_1}' d_2 - ia' (\bar{d_2}d_1 + \bar{d_1} d_2))]A =\delta^{\mu}_{0}[(-Q^A) (-A)]\nonumber\\
J^{\mu}_{C} &= \delta^{\mu}_{0}\frac{k}{4\pi}Tr[2i a' \mathcal{C}] = \delta^{\mu}_{0} [Q_C (-i \mathcal{C})], \quad Q_C = -\frac{k a'}{2 \pi}\nonumber\\
J^{\mu}_{\Theta} &= \delta^{\mu}_{0}\frac{k}{2\pi} Tr[\{\lambda^{-1}\hat\alpha \lambda + 2u (\lambda^{-1}\lambda')'- \mu \lambda^{-1}\lambda' \}\theta]=  \delta^{\mu}_{0} 2 Tr[Q^J_a \Theta^a]\nonumber\\
J^{\mu}_{N} &= \delta^{\mu}_{0} \frac{k}{4\pi}Tr[-4\lambda^{-1}\lambda' N] = \delta^{\mu}_{0} 2 Tr[Q^P_a (-N^a)]\\
J^{\mu}_{D_2} &= \delta^{\mu}_{0}(-\frac{k}{\pi})Tr [(\bar{d_1}' \lambda + i a' \bar{d_1} \lambda) D_2] = Tr[Q^{G_2}_{\alpha}D_2^{\alpha}]\nonumber\\
J^{\mu}_{D_1} &= \delta^{\mu}_{0}(-\frac{k}{\pi}) Tr[(\bar{d_2}' \lambda - i a' \bar{d_2} \lambda) D_1]= Tr[Q^{G_1}_{\alpha}D_1^{\alpha}],\nonumber
\end{align}
where $N(\phi)$ and $\Theta(\phi)$ are infinitesimal $SL(2,\mathbb{R})$ matrices which can be further expanded in the basis of $\Gamma$ matrices as $N(\phi) = N^a(\phi) \Gamma_a$ and $\Theta(\phi) = \Theta^a(\phi) \Gamma_a$.
The details of the above computations can be found in appendix \ref{AppD}. Here we have chosen $S^{[\mu\nu]}_{\epsilon}, T^{\mu}_{\epsilon}$ for certain transformations, as we want the current to be non zero only along the time $u$ component. This way we directly get the canonical generators of the corresponding transformation.  From these currents, one can find the corresponding current algebra using the usual procedure\cite{1974AnPhy..88..286R}. Alternate way to get to the same algebra is, in Hamiltonian formalism, the computation of the Dirac bracket algebra of the canonical generators of the symmetries using the relation below, 
$$ \delta_{\epsilon_2}J^0_{\epsilon_1}= \{J^0_{\epsilon_1},J^0_{\epsilon_2}\}_{DB}
$$
The Dirac brackets calculated are given below: 
\begin{align}\label{CA}
\left\lbrace  Q^P_a(\phi), Q^P_b(\phi')    \right\rbrace_{DB} &= \left\lbrace  Q^P_a(\phi), Q^A(\phi')    \right\rbrace_{DB}=\left\lbrace   Q^P_a(\phi), Q^C(\phi')    \right\rbrace_{DB}=0  \nonumber\\
\left\lbrace  Q^P_a(\phi),  Q^{G_1}_{\alpha}(\phi')    \right\rbrace_{DB} &=\left\lbrace  Q^P_a(\phi), Q^{G_2}_{\alpha}(\phi')    \right\rbrace_{DB} =0 \nonumber\\
\left\lbrace  Q^P_a(\phi), Q^J_b(\phi')    \right\rbrace_{DB} &= \left\lbrace  Q^J_a(\phi), Q^P_b(\phi')    \right\rbrace_{DB} =\epsilon_{abc} Q^P_c(\phi) \delta(\phi - \phi')- \frac{k}{2\pi} \eta_{ab} \partial_{\phi}\delta(\phi-\phi')\nonumber\\
\left\lbrace  Q^J_a(\phi), Q^J_b(\phi')    \right\rbrace_{DB} &= \epsilon_{abc} Q^J_c(\phi) \delta(\phi - \phi')+\mu \frac{k}{2\pi} \eta_{ab} \partial_{\phi}\delta(\phi-\phi')\nonumber\\
\left\lbrace  Q^J_a(\phi), Q^A(\phi')    \right\rbrace_{DB} &=\left\lbrace   Q^J_a(\phi), Q^C(\phi')    \right\rbrace_{DB}=0\\
\left\lbrace  Q^{G_1}_{\alpha}(\phi), Q^J_a(\phi')    \right\rbrace_{DB} &=-\frac{1}{2} (\Gamma_a)^{\beta}_{\alpha} Q^{G_1}_{\beta}(\phi) \delta(\phi-\phi')\nonumber\\
\left\lbrace  Q^{G_2}_{\alpha}(\phi), Q^J_a(\phi')    \right\rbrace_{DB} &=-\frac{1}{2} (\Gamma_a)^{\beta}_{\alpha} Q^{G_2}_{\beta}(\phi) \delta(\phi-\phi')\nonumber\\
\left\lbrace  Q^C(\phi), Q^C(\phi')    \right\rbrace_{DB} &=\left\lbrace  Q^C(\phi),  Q^{G_1}_{\alpha}(\phi')    \right\rbrace_{DB} =\left\lbrace  Q^C(\phi), Q^{G_2}_{\alpha}(\phi')    \right\rbrace_{DB} =0\nonumber\\
\left\lbrace  Q^C(\phi), Q^A(\phi')    \right\rbrace_{DB} &= \frac{k}{2\pi } \partial_{\phi} \delta(\phi-\phi'),\quad
\left\lbrace  Q^{A}(\phi), Q^{A}(\phi')    \right\rbrace_{DB} = \frac{k}{2\pi} \bar{\mu} \partial_{\phi} \delta(\phi-\phi')\nonumber\\
\left\lbrace  Q^{G_1}_{\alpha}(\phi), Q^{A}(\phi')    \right\rbrace_{DB}  &=- i Q^{G_1}_{\alpha}(\phi) \delta(\phi - \phi'), \quad
\left\lbrace  Q^{G_2}_{\alpha}(\phi), Q^{A}(\phi')    \right\rbrace_{DB} = i Q^{G_2}_{\alpha}(\phi) \delta(\phi - \phi')\nonumber\\
\left\lbrace  Q^{G_1}_{\alpha}(\phi), Q^{G_2}_{\beta}(\phi')   \right\rbrace_{DB} &= -(C\Gamma^a)_{\alpha\beta} Q^P_a \delta(\phi-\phi') - \frac{k}{\pi} C_{\alpha \beta} \partial_{\phi} \delta(\phi-\phi') +ia' \frac{k}{\pi} C_{\alpha\beta} \delta(\phi-\phi'),\nonumber
\end{align}
here we have used  $\frac{k}{\pi} C_{\alpha \beta} (\lambda^{-1}\lambda')^{\beta}_{\gamma} = (C\Gamma^a)_{\alpha \gamma} Tr[\Gamma_a (\lambda^{-1}\lambda')]$.
This is the same affine extended ${\cal{N}}=(2,0)$ SuperPoincar\'{e} algebra after a change of basis for the fermionic generators as,
\begin{align*}
Q^{1}_{\alpha}(\phi) = \frac{1}{2} (Q^{G_1}_{\alpha}(\phi) + Q^{G_2}_{\alpha}(\phi)), \quad
Q^{2}_{\alpha} (\phi) = \frac{1}{2i} (Q^{G_1}_{\alpha}(\phi) - Q^{G_2}_{\alpha}(\phi)).
\end{align*}
In this new basis the fermionic Dirac Brackets take the form:
\begin{align*}
\lbrace Q^1_{\alpha}(\phi), Q^2_{\beta}(\phi')   \rbrace_{DB} &= -\frac{k}{2\pi} a'(\phi) C_{\alpha \beta} \delta'(\phi -\phi')\\
\lbrace Q^I_{\alpha}(\phi), Q^I_{\beta}(\phi')   \rbrace_{DB} &= -\frac{1}{2} (C \Gamma^a)_{\alpha \beta} Q^P_a(\phi) \delta(\phi -\phi') - \frac{k}{2\pi} C_{\alpha \beta} \delta'(\phi -\phi')\\
\lbrace  Q^{1}_{\alpha}(\phi), Q^{A}(\phi') \rbrace_{DB} &= Q^{2}_{\alpha}(\phi) \delta(\phi - \phi'), \quad
\lbrace  Q^{2}_{\alpha}(\phi), Q^{A}(\phi') \rbrace_{DB} = -Q^{1}_{\alpha}(\phi) \delta(\phi - \phi')\\
\left\lbrace  Q^J_a(\phi), Q^{1}_{\alpha}(\phi')    \right\rbrace_{DB} &=\frac{1}{2} (\Gamma_a)^{\beta}_{\alpha} Q^{1}_{\beta}(\phi) \delta(\phi-\phi'), \quad
\left\lbrace   Q^J_a(\phi),Q^{2}_{\alpha}(\phi')    \right\rbrace_{DB} =\frac{1}{2} (\Gamma_a)^{\beta}_{\alpha} Q^{2}_{\beta}(\phi) \delta(\phi-\phi').
\end{align*}
The above modified Dirac brackets along with bosonic ones in \eqref{CA} reproduce the exact affine extended ${\cal{N}}=(2,0)$ SuperPoincar\'{e} algebra that we started with in \eqref{N=(2,0)}.
 Thus, we see that the global symmetry of the chiral WZW theory is exactly same that of the dual 3D Supergravity. Similarly for ${\cal{N}}=(1,1)$ case we get following affine extension,
\begin{eqnarray*}
\label{PP}
\{Q^P_{a}(\phi),Q^P_{a}(\phi')\}_{DB}&=&0\\
\{Q^P_{a}(\phi),Q^J_{b}(\phi')\}_{DB}&=&\epsilon_{abc}Q^{P}_{c}(\phi)\delta(\phi-\phi')-\frac{k}{2\pi}\eta_{ab}\delta'(\phi-\phi')\\
\{Q^P_{a}(\phi),Q^{i}_{\alpha}(\phi')\}_{DB}&=&0 \quad (i=1,2)\\
\{Q^J_{a}(\phi),J_{b}(\phi')\}_{DB}&=&\epsilon_{abc}Q^{J}_{c}\delta(\phi-\phi')+\frac{\mu k}{2\pi}\eta_{ab}\delta'(\phi-\phi') \\
\{Q^J_{a}(\phi),Q_{\alpha}^{i}(\phi')\}_{DB}&=&\half(Q^{i}\Gamma_{a})_{\alpha}(\phi)\delta(\phi-\phi')\quad (i=1,2)\\
\label{QQ}
\{Q_{\alpha}^{i}(\phi),Q_{\beta}^{j}(\phi')\}_{DB}&=&\delta^{ij}[-\half(C\Gamma^{a})_{\alpha\beta}Q^P_{a}(\phi)\delta(\phi-\phi')-\frac{k}{2\pi}C_{\alpha\beta}\delta'(\phi-\phi')]
\end{eqnarray*}
The explicit derivation of various canonical generators for ${\cal{N}}=(1,1)$ case has been provided in appendix \ref{AppC}. 
\subsection{Gauge Symmetry}\label{sec5.2}
Other than the above global symmetry, the chairal WZW model \eqref{wzwa2} is also invariant under a gauge symmetry. The gauge transformations of various fields can be obtained from the Polyakov-Wiegmann identities and for an arbitrary gauge transformation parameter $\Sigma (u)$, the transformations are given as follows :
\begin{equation}\label{gt}
\lambda (u) \rightarrow \Sigma (u) \lambda , \quad d_i (u) \rightarrow \Sigma (u)  d_i (u) \ \ \  i=1,2, \quad F(u) \rightarrow \Sigma (u) F \Sigma (u)^{-1},
\end{equation}
while $a(u), c(u)$ remains invariant. This makes the dynamics of this system constrained and we need to take into account its implications in defining the conserved charges of the theory. We shall come back to this issue in the next section.

\section{Enhanced Symmetries of $\mathcal{N}=2$ SuperPoincar\'{e} Wess-Zumino-Witten theory}\label{sec6}
In order to get an infinite dimensional mode algebra from the above current algebra usual conformal field theory techniques of \cite{DiFrancesco:639405} can be used after a slight modification.  We implement the modified Sugawara construction following \cite{Barnich:2015sca} to get the stress-tensor. In this case, we are looking for four bosonic generators and two fermionic generators. These can be achieved by defining the followings:
\begin{align}\label{ag}
H &= \frac{\pi}{k} Q^P_a Q^P_a +4 \frac{\pi}{k} Q^C Q^C \nonumber\\
J &= -\mu \frac{\pi}{k} Q^P_a Q^P_a - 2 \frac{\pi}{k} Q^J_a Q^P_a + \frac{\pi}{k} C_{\alpha \beta} Q^{G_1}_{\alpha} Q^{G_2}_{\beta}+ 2 \frac{\pi}{k} Q^A Q^C- \bar{\mu} \frac{\pi}{k} Q^C Q^C\\
\mathcal{G}^1 &= \frac{\pi}{k} (Q^P_2 Q^{G_1}_+ +\sqrt{2} Q^P_0 Q^{G_1}_-)+2i \frac{\pi}{k}  Q^{G_1}_+ Q^C, \nonumber \\
\mathcal{G}^2 &=\frac{\pi}{k} (Q^P_2 Q^{G_2}_+ +\sqrt{2} Q^P_0 Q^{G_2}_-)- 2 i \frac{\pi}{k}  Q^{G_1}_+ Q^C \nonumber
\end{align}
along with $Q^A$ and $Q^C$ defined in the last section. 
 Here $H,J$ are both weight two bosonic generators and $J$ corresponds to the stress-tensor. $Q^A, Q^C$ are two weight one bosonic generators and $\mathcal{G}^1,\mathcal{G}^2$ are two weight $3/2$ fermionic generators. The values of the relative coefficients in the bilinear of currents are fixed by demanding that the Dirac brackets of stress mode $J$ with other bosonic generators should be proportional to them  i.e. $\lbrace J(\phi), Q(\phi') \rbrace\sim Q(\phi)$ for each current mode $Q(\phi)$. We refer the readers to appendix E for computational details. 

There is a subtle point to note here. The chiral WZW model that we are studying is a gauge theory. Thus, in the usual Hamiltonian formulation, we must study it as a constrained system. We refer the reader to \cite{Henneaux:1992ig} for a detailed review on constrained systems. The constraints\footnote{a primary constraint is a relation among canonical variables that needs to be satisfied, without imposing eom.} arise from the imposed  boundary (asymptotic) value of the radial gauge fixed Chern-Simons field, $a = g^{-1}dg$, as given in \eqref{bgf}. We have already taken into account part of the constraints (type (a)) for constructing the corresponding WZW model. The remaining constraints(type (b)) on the gauge field parameters are 
\begin{align*}
\hat\omega^1_{\phi}=\sqrt{2}; \quad \omega^2_{\phi}= 0; \quad \psi^{1+}_{\phi}= \psi^{2+}_{\phi} = 0; \quad e^1_{\phi}=e^2_{\phi}=0.
\end{align*}
These conditions, that need to be imposed only at the boundary, manifest themselves through constraints on the fields of the WZW model \eqref{gfa}. This is because, at the boundary we can as well identify the onshell CS gauge field parameters $\hat\omega, \psi^1,\psi^2, e$ of \eqref{wb},\eqref{g1g2},\eqref{ce}  with the WZW fields $\lambda,  d_1,d_2,\hat\alpha$ of \eqref{gff}\footnote{at the boundary, the non dynamical functions (of $u,r$) can get absorbed into the WZW fields.}. 
Thus the constraints on fields are:
\begin{align*}
(\lambda^{-1}\lambda')^1 = \sqrt{2} \qquad (\lambda^{-1}\lambda')^2=0\\
(\lambda^{-1} d'_1 + ia' \lambda^{-1} d_1)^- =
(\lambda^{-1} d'_2 - ia' \lambda^{-1} d_2)^- =0 \\
(\lambda^{-1}\frac{\hat\alpha}{2}\lambda)^1 = (\lambda^{-1}\frac{\hat\alpha}{2}\lambda)^2=0.
\end{align*}
The above constraints can as well be expressed in terms of the canonical generators of \eqref{cc}as, 
\begin{align}\label{2cc}
Q^P_0 &= \sqrt{2} \frac{k}{2\pi}, \quad Q^P_2 =0 \nonumber\\
Q^J_0 &= -\sqrt{2} \frac{\mu k}{2\pi}, \quad Q^J_2 =0\\
Q^1_{1+}&=0, \qquad Q^2_{+}=0\nonumber
\end{align}
Let us denote the above constraints as $\{\Phi_l\}, l=1, \cdots 6$ respectively as presented above. They collectively define the constrained hypersurface. It can be easily verified that, four out of these six constraints, denoted as $\{\gamma_p\}=\{\Phi_1,\Phi_3,\Phi_4,\Phi_6\}, p=1,\cdots 4$ have a vanishing Dirac brackets with all of  $\{\Phi_l\}$ on the constrained hypersurface
\footnote{$
\lbrace \phi_2,\phi_4 \rbrace =\lbrace Q^P_2, Q^J_2 \rbrace  \neq 0$, as there is a central term.}. Thus $\{\gamma_p\}$ are the firstclass constraints and they generate the gauge symmetry that we presented in subsection \ref{sec5.2}. It is a well known fact\cite{Henneaux:1992ig} that in a constrained system, one needs to usually modify the canonical generators (conserved charges) such that they commute with the first class constraints on the constraint hypersurface (defined by first class ones) as otherwise they are not gauge invariant (and hence are not physical observable). The 
charges defined in\eqref{ag} fail to satisfy this property, as we have shown in appendix \ref{AppE}. Thus we need to further modify them using \eqref{nci}, such that the resultant charges are true observables of the theory.  
The required minimal shifts in generators that achieve the above requirements are given by:
\begin{align*}
{\cal{H}} = H + \partial_{\phi} Q^P_2; \hspace{15pt} {\cal{J}} = J - \partial_{\phi} Q^J_2; \hspace{15pt} \hat{\mathcal{G}}^I= \mathcal{G}^I + \partial_{\phi} Q^I_{+}
\end{align*}
Finally we compute the Dirac brackets of these new gauge invariant canonical generators and they are given as\footnote{look at appendix \ref{AppF}},
\begin{align}\label{n20pb}
\lbrace {\cal{J}}(\phi),  {\cal{J}}(\phi') \rbrace _{DB}&= ( {\cal{J}}(\phi) +  {\cal{J}}(\phi')) \partial_{\phi} \delta(\phi - \phi') -\mu  \frac{k}{2\pi} \partial_{\phi}^3 \delta(\phi-\phi')\nonumber\\
\lbrace {\cal{H}}(\phi),  {\cal{J}}(\phi') \rbrace_{DB} &= ({\cal{H}}(\phi) + {\cal{H}}(\phi')) \partial_{\phi} \delta(\phi - \phi')- \frac{k}{2\pi} \partial_{\phi}^3 \delta(\phi-\phi')\nonumber\\
\{\tilde{\mathcal{H}}(\phi),\tilde{\mathcal{H}}(\phi')\}_{DB}&=0, \quad \lbrace{\cal{H}}(\phi), Q^A(\phi') \rbrace_{DB} =4 Q^C(\phi) \partial_{\phi} \delta(\phi - \phi')\nonumber\\
\lbrace  {\cal{J}}(\phi), Q^A(\phi') \rbrace_{DB} &=  Q^A(\phi) \partial_{\phi}\delta(\phi-\phi'), \quad \lbrace  {\cal{J}}(\phi), Q^C(\phi') \rbrace_{DB} = Q^C(\phi)\partial_{\phi} \delta(\phi - \phi')\nonumber\\
\lbrace  {\cal{J}}(\phi), Q^A(\phi') \rbrace_{DB} &=  Q^A(\phi) \partial_{\phi}\delta(\phi-\phi')\\
\left\lbrace  Q^C(\phi), Q_A(\phi')    \right\rbrace_{DB} &=  \frac{k}{2\pi} \partial_{\phi} \delta(\phi-\phi'), \quad \left\lbrace  Q^A(\phi), Q^A(\phi')    \right\rbrace_{DB} =  \frac{k}{2\pi} \bar \mu\partial_{\phi} \delta(\phi-\phi')\nonumber\\
\lbrace  {\cal{J}}(\phi), \hat {\mathcal{G}}^i(\phi') \rbrace_{DB} &= (\hat {\mathcal{G}}^i(\phi) +\frac{1}{2} \hat {\mathcal{G}}^i(\phi'))\partial_{\phi} \delta(\phi - \phi'), \quad (i=1,2)\nonumber\\
\lbrace  {\cal{H}}(\phi), \hat {\mathcal{G}}^i(\phi') \rbrace_{DB} &= 0, \quad (i=1,2)\nonumber\\
\lbrace  \hat {\mathcal{G}}^1(\phi),Q^A(\phi')  \rbrace_{DB} &= -i  \hat {\mathcal{G}}^1(\phi) \delta (\phi-\phi'), \quad
\lbrace \hat {\mathcal{G}}^2(\phi),Q^A(\phi')  \rbrace_{DB} = i \hat {\mathcal{G}}^2(\phi) \delta (\phi-\phi')\nonumber\\
\lbrace \hat {\mathcal{G}}^1(\phi),\hat {\mathcal{G}}^2(\phi')  \rbrace_{DB} &=  {\cal{H}}(\phi)\delta(\phi-\phi')-\frac{k}{\pi}\partial^2_{\phi}\delta(\phi-\phi') -2i(Q^C(\phi)+Q^C(\phi'))\delta'(\phi-\phi') \nonumber
\end{align}
The above Dirac brackets are expected to be the ones of the physical observables ${\cal{H}} , {\cal{J}},\hat{\mathcal{G}}^i,Q^A,Q^C$ of the reduced two dimensional Super Liouville theory that is dynamically equivalent to the 3D Supergravity. We shall report on the details structure of the Liouville theory in a separate work \cite{WIP}.

Let us also present the enhanced symmetry for the ${\cal{N}}=(1,1)$ case here :
\begin{eqnarray}\label{n11pb}
\{{\mathcal{H}}(\phi),{\mathcal{H}}(\phi')\}_{DB}&=&0\nonumber\\
\{{\mathcal{H}}(\phi),{\mathcal{J}}(\phi')\}_{DB}&=&({\mathcal{H}}(\phi)+{\mathcal{H}}(\phi'))\delta'(\phi-\phi')-\frac{k}{2\pi}\partial^{3}\delta(\phi-\phi')\nonumber\\
\{{\mathcal{H}}(\phi),{\mathcal{G}}^{i}(\phi')\}_{DB}&=&0\\
\{{\mathcal{J}}(\phi),\tilde{\mathcal{J}}(\phi')\}_{DB}&=&({\mathcal{J}}(\phi)+{\mathcal{J}}(\phi'))\delta'(\phi-\phi')-\frac{\mu k}{2\pi}\partial^{3}\delta(\phi-\phi')\nonumber\\
\{{\mathcal{J}}(\phi),\tilde{\mathcal{G}}^{i}(\phi')\}_{DB}&=&(\tilde{\mathcal{G}}^{i}(\phi)+\half\tilde{\mathcal{G}}^{i}(\phi'))\delta'(\phi-\phi')\nonumber\\
\{\tilde{\mathcal{G}}^{i}(\phi),\tilde{\mathcal{G}}^{i}(\phi')\}_{DB}&=&\delta^{ij}(\tilde{\mathcal{H}}(\phi)\delta(\phi-\phi')-\frac{k}{2\pi}\partial^{2}\delta(\phi-\phi')).\nonumber
\end{eqnarray}
Here the physical observables ${\cal{H}} , {\cal{J}},\tilde{\mathcal{G}}^i$ of the reduced two dimensional Super Liouville theory is dynamically equivalent to a 3D Supergravity with two supercharges but without any internal R-symmetry.

\section{A new ${\cal{N}}=2$ SuperBMS$_3$ algebra}\label{sec7}

Finally we write the quantum algebra that corresponds to the above Dirac brackets \eqref{n11pb} and \eqref{n20pb}. For this purpose,  we define 
The modes of the above fields as.
\begin{align*}
M_n &= \frac{1}{2 \pi}\int d\phi e^{in\phi}{\cal{H}}(\phi), \quad 
J_n = \frac{1}{2 \pi}\int d\phi e^{in\phi}{\cal{J}}(\phi),\\
\mathcal{G}^{1,2}_n &= \frac{1}{2 \pi}\int d\phi e^{in\phi}\hat{\mathcal{G}}^{1,2}(\phi), \quad \tilde{\mathcal{G}}^{1,2}_n = \frac{1}{2 \pi}\int d\phi e^{in\phi}\tilde{\mathcal{G}}^{1,2}(\phi),\\
R_n &= \frac{1}{2 \pi}\int d\phi e^{in\phi}Q^A(\phi), \quad
S_n = \frac{1}{2 \pi}\int d\phi e^{in\phi}Q^C(\phi).
\end{align*}
We further use the identification for bosonic and fermionic commutator brackets respectively as $$i \{,\}_{DB} \rightarrow  [,] \qquad \{,\}_{DB} \rightarrow \{,\}.$$
The non zero brackets of the algebra corresponding to \eqref{n20pb} looks as,
\begin{align}
[M_n,J_m] &=(n-m)M_{n+m}+n^3k\;\delta_{n+m,0}, \quad 
[J_n,J_m] =(n-m)J_{n+m}+n^3\mu\;k\;\delta_{n+m,0}\nonumber\\
[M_n,R_m]&= - 4m S_{n+m}, \quad [J_n,R_m]  =-m R_{n+m} , \quad [J_n,S_m] = - m S_{n+m}\nonumber\\
[R_n, S_m]&=n\;k\;\delta_{n+m,0}, \quad [R_n, R_m]= n \ \bar \mu \;k\;\delta_{n+m,0}\nonumber\\
[J_n, \mathcal{G}^i_m]&=\;\left(\frac{n}{2}-m \right)\;\mathcal{G}^i_{n+m}, \quad (i=1,2)\\
[R_n, \mathcal{G}^1_m]&=\mathcal{G}^1_{n+m}, \quad [R_n, \mathcal{G}^2_m]=-\mathcal{G}^2_{n+m} \nonumber\\
\{\mathcal{G}^{1}_n,\mathcal{G}^2_m\} &= M_{n+m}+ 2 k n^2 \delta_{n+m,0}+(n-m)S_{n+m}\nonumber
\end{align}

This is a a new SuperBMS$_3$ algebra, so far not identified in the literature. Here the central term for $[J_n,J_m] $ and $[R_n, R_m]$ are independent of each other. 
The closest one that was formulated in \cite{Caroca:2018obf,Basu:2017aqn} has both these central terms identical and the other one obtained in \cite{Fuentealba:2017fck} has zero central extension for both commutators. We see that the 2D dual theory constructed in \eqref{gfa} has a richer quantum symmetry. 

A similar analysis form \eqref{n11pb} reproduces the ${\cal{N}}=2$ SuperBMS$_3$ algebra that was introduced in \cite{Banerjee:2016nio}.

\section{Outlook}\label{sec8}
In this paper, we have found the most generic field theory dual to 3-dimensional asymptotically flat ${\cal{N}}=2$ Supergravity. We have constructed the duals for both  ${\cal{N}}=(1,1)$ and ${\cal{N}}=(2,0)$ cases. The physical observables dual to 3D graviton (and other supergravity fields) belong to a super Liouville like theory. We found that for ${\cal{N}}=(2,0)$ case the dual theory enjoys an infinite dimensional most generic ${\cal{N}}=2$ quantum SuperBMS$_3$ symmetry that so far was not known. The symmetry is a truncated version of ${\cal{N}}=4$ quantum SuperBMS$_3$ that we developed in \cite{Banerjee:2016nio}. The most interesting feature we noticed in \cite{Banerjee:2016nio}is that, in presence of $R-$symmetry, independent central extensions are possible for $[J_n,J_m] $ and $[R_n, R_m]$ commutators. This is supported by Jacobi identity and the free field realisation of the algebra. Here we find that the same noble feature is true even for ${\cal{N}}=2$ SuperBMS$_3$ algebra.

The phase space dynamics of the reduced WZW model is governed by a generalised Liouville theory.
 We shall report in details on the dual Liouville like theory in a future work. More over it would be interesting to see how this theory can provide a microscopic understanding of the cosmological solutions found in \cite{Fuentealba:2017fck}.

Three dimensional gravity does not have a propagating graviton. Thus all its non trivial dynamics are only governed by the boundary degrees of freedom. Our theory (and \cite{Barnich:2013yka,Barnich:2015sca} for simpler cases) is the one that describes this dynamics. It would be nice to see how one can use these theories to answer interesting physics of 3D gravity.

Earlier in \cite{Banerjee:2015kcx,Banerjee:2016nio} a free field realisation of SuperBMS$_3$ algebras was presented. It would be interesting to see how the theory constructed in this paper are related to those.  
Another technically challenging problem would be to extend the above analysis for ${\cal{N}}=4$ \cite{Banerjee:2017gzj,Ito:1998vd} and ${\cal{N}}=8$ \cite{Marcus:1983hb,Banerjee:2018hbl} Supergravity theories. With that, we shall have a complete zoo of all 2D duals of all possible 3D Supergravities.

\vspace{1cm}
{\bf Acknowledgements}\\

We would like to thank Glenn Barnich for a very useful communication at an important stage of this work. We acknowledge useful discussions with Javier Matulich. NB acknowledges hospitality at ICTP during the final stage of this work. Our work is partially supported by a SERB ECR grant, GOVT of India. AB would like to thank IISER Bhopal for their hospitality throughout the major part of this project. For TN, the work is partially supported by the ERC Advanced Grant “High-Spin-Grav” and by FNRS-Belgium (convention IISN 4.4503.15). Finally, we thank the people of India for their generous support for the basic sciences.

\appendix

\section{Conventions and Identities}\label{AppA}

In this paper, we have mostly followed the conventions of \cite{Barnich:2015sca}. The tangent space metric $\eta_{ab}, a=0,1,2$ is flat and off-diagonal, given as
\begin{equation}
\nonumber
\eta_{ab}=\left(\begin{matrix}
0&1&0\\
1&0&0\\
0&0&1\\
\end{matrix}\right) .
\end{equation}
The space time coordinates are $u, \phi,r$ with positive orientation in the bulk being $du d\phi dr$. Accordingly the Levi-Civita symbol is chosen such that 
$\epsilon_{012}=1$.\\
The three dimensional Dirac matrices satisfy usual commutation relation  
$\{\Gamma_{a},\Gamma_{b}\}=2\eta_{ab}$ . They also satisfy following useful identities:
$$\Gamma_{a}\Gamma_{b}=\epsilon_{abc}\Gamma^{c}+\eta_{ab}\mathbb{I}, \hspace{21pt} (\Gamma^{a})^{\alpha}_{\beta}(\Gamma_{a})^{\gamma}_{\delta}=2\delta^{\alpha}_{\delta}\delta^{\gamma}_{\beta}-\delta^{\alpha}_{\beta}\delta^{\gamma}_{\delta}.$$
The explicit form of the Dirac matrices are chosen as,
\begin{equation}
\Gamma_0 = \sqrt{2}\left(\begin{array}{cc} 0 & 1 \\ 0 & 0\end{array} \right)\,, \qquad 
\Gamma_1 =\sqrt{2} \left(\begin{array}{cc} 0 & 0 \\ 1 & 0\end{array} \right)\,, \qquad
\Gamma_2 = \left(\begin{array}{cc} 1 & 0 \\ 0 & -1\end{array} \right)\,.
\end{equation} 
All spinors in this work are Majorana and our convention for the majorana conjugate of the fermions are different from \cite{Barnich:2015sca} and is given as, 
$$ \overline{\psi}_{\alpha i}=\psi^{\beta}_iC_{\beta\alpha}, \qquad
C_{\alpha\beta}=\epsilon_{\alpha\beta}=C^{\alpha\beta}=\left(\begin{matrix}
0&1\\
-1&0\\
\end{matrix}\right).\\ $$
Here $i=1,2$ is the internal index and $C_{\alpha\beta}$ is the charge conjugation matrix that satisfies 
$$C^{T}=-C, \hspace{11pt} C\Gamma_{a}C^{-1}=-(\Gamma_{a})^{T},\hspace{11pt} C_{\alpha \beta} C_{\beta \gamma} = - \delta_{\alpha \gamma} $$
For any traceless $2 \times 2$ matrix $A$, it can be shown that $C_{\alpha \beta} A^{\beta}_{\gamma} = (C\Gamma^a)_{\alpha\gamma} Tr[\Gamma_a A]$.\\
For computing the gauged action the three dimensional Fierz relation is useful and is given as
\begin{equation}\label{Fierz}
\zeta\bar{\eta} = - \frac12 \bar{\eta}\, \zeta \, \mathbbm{1} - \frac12 (\bar{\eta}\Gamma^a \zeta)\Gamma_a\;,
\end{equation}

Other useful identities are:

\begin{equation}
\bar\psi \G_a\,\eta=\bar\eta\,\G_a\,\psi\;,\qquad \bar\psi \G_a\,\epsilon= -\bar\epsilon\,\G_a\,\psi\;,
\end{equation}
where $\psi,\eta$ are Grassmannian one-forms, while 
$\epsilon$ is a Grassmann parameter.

\section{Hamiltonian form of the CS action}\label{AppB}
In this appendix, we shall present the details of the Hamiltonian action and the boundary term corresponding to a Chern-Simons theory on a 3 manifold with boundary.
 We decompose the gauge field $A(u,\phi,r)$ as $A=duA_{u}+\tilde{A}$. Thus we give a preference to the time like $u$ direction and other two directions are treated together. The reasoning behind this decomposition is: in variational principle, in general we can not through out the variations of derivatives of gauge fields along the spacelike directions. Next we can decompose the field strength.
Using $A=du\;A_u+\tilde{A}$ and $d=du\;\partial_u+\tilde{d}$, we get 
\begin{equation}
dA=du \dot{\tilde{A}} +\tilde{d} \;duA_u+\tilde{d}\tilde{A}
\end{equation}
Therefore
\begin{align*}
&<A,dA> \\
&=<(du\;A_u+\tilde{A}),(du \dot{\tilde{A}} +\tilde{d} \;duA_u+\tilde{d}\tilde{A} )> \\
&=<du\;A_u,\tilde{d}\tilde{A}>+<\tilde{A}, du\dot{\tilde{A}}>+<\tilde{A},\tilde{d}duA_u> \\
&=<\tilde{A}, du\dot{\tilde{A}}>+2 <du\;A_u,\tilde{d}\tilde{A}> +\text{total derivative term}
\end{align*}
where we have used 
\begin{align*}
\tilde{d}<\tilde{A},duA_u> &=<\tilde{d}\tilde{A},duA_u> - <\tilde{A},\tilde{d}duA_u> \\
&=<,duA_u,\tilde{d}\tilde{A}> - <\tilde{A},\tilde{d}duA_u> 
\end{align*}
using cyclic invariance of trace in the last step.
~~\\
Also we have, $(A\wedge A)=(duA_u\wedge\tilde{A}+\tilde{A}\wedge duA_u+\tilde{A}\wedge\tilde{A}).$
Therefore
\begin{align*}
<A^3>&=<(du\;A_u+\tilde{A})\wedge(duA_u\wedge\tilde{A}+\tilde{A}\wedge duA_u+\tilde{A}\wedge\tilde{A})> \\
&=<duA_u\wedge\tilde{A}\wedge\tilde{A}>+<\tilde{A}\wedge duA_u\wedge\tilde{A}>+<\tilde{A}\wedge\tilde{A}\wedge duA_u> \\
&=3<duA_u\wedge\tilde{A}\wedge\tilde{A}>
\end{align*}
Now collecting all the terms 
and putting in the CS action \eqref{csaction} we get,
\begin{align}\label{HCS}
I_H[A]=\frac{k}{4\pi} \int <\tilde{A},du\dot{\tilde{A}}>+2<duA_u,\tilde{d}\tilde{A}+\tilde{A}^2>
\end{align}
Finally we present construction of the boundary term.
Variation of the above Hamiltonian form of the Chern-Simons action \eqref{HCS}s is given by
\begin{eqnarray}
\label{actionVariation}
\delta I_{H}[A]&=& \frac{k}{4\pi}\int \delta\langle \tilde{A},\dot{\tilde{A}}du \rangle + 2\langle du \delta A_{u},\tilde{d}\tilde{A}+\tilde{A}^{2}\rangle + 2\langle du A_{u},\tilde{d}\delta\tilde{A}+\delta\tilde{A}^{2}\rangle\\
\nonumber\\
\nonumber
\delta\langle \tilde{A},\dot{\tilde{A}}du \rangle&=&dud\phi dr\hspace{3pt} Tr[-\delta A_{r}\partial_{u}A_{\phi}+\delta A_{\phi}\partial_{u}A_{r}-A_{r}\partial_{u}(\delta A_{\phi})+A_{\phi}\partial_{u}(\delta A_{r})]\\
\nonumber
\tilde{d}\tilde{A}&=&d\phi dr(\partial_{r}A_{\phi}-\partial_{\phi}A_{r}), \quad \tilde{A}^{2}=d\phi dr(A_{\phi}A_{r}-A_{r}A_{\phi}), \quad \tilde{d}\tilde{A}+\tilde{A}^{2}=d\phi dr \mathcal{F}_{r\phi}
\end{eqnarray}
Substituting all these expressions in (\ref{actionVariation}), we get
\begin{eqnarray}
\nonumber
\delta I_{H}[A]&=&\frac{k}{4\pi}\int du d\phi dr Tr[-\delta A_{r}\partial_{u}A_{\phi}+\delta A_{\phi}\partial_{u}A_{r}-{\color{blue}A_{r}\partial_{u}(\delta A_{\phi})}+{\color{blue}A_{\phi}\partial_{u}(\delta A_{r})}]+\frac{k}{2\pi}\int du d\phi dr Tr[\delta A_{u}\mathcal{F}_{r\phi}]\\
\nonumber
&&+\frac{k}{2\pi}\int du d\phi dr Tr[{\color{red}A_{u}\partial_{r}(\delta A_{\phi})}-{\color{blue}A_{u}\partial_{\phi}(\delta A_{r})}+A_{u}(\delta A_{\phi}A_{r}+A_{\phi}\delta A_{r}-\delta A_{r}A_{\phi}-A_{r}\delta A_{\phi})]
\end{eqnarray}
The colored terms can be manipulated to write them as a total derivative plus another term. The total derivative terms from all the blue terms can be integrated out to give zero at the boundary (as the variation of fields are zero at the boundary). The red colored term gives a non-zero term at the boundary $r=r_{0}$ (marked green in the following expression). All other terms combine to give the following variation of the action:
\begin{eqnarray}
\nonumber
\delta I_{H}[A]&=&\frac{k}{2\pi}\int du d\phi dr Tr[\delta A_{u}\mathcal{F}_{r\phi}]+\frac{k}{2\pi}\int du d\phi dr Tr[\delta A_{
	\phi}\mathcal{F}_{u r}]+\frac{k}{2\pi}\int du d\phi dr Tr[\delta A_{r}\mathcal{F}_{\phi u}]\\
&&{\color{teal}+\frac{k}{2\pi}\int du d\phi dr Tr[\partial_{r}(A_{u}\delta A_{\phi})]-\frac{k}{2\pi}\int du d\phi dr Tr[\partial_{\phi}(A_{u}\delta A_{r})]}
\end{eqnarray}
The boundary term can be rewritten in form notation as $-\frac{k}{2\pi}\int du\tilde{d}\langle A_{u},\delta\tilde{A}\rangle $ as,
\begin{equation}
-\frac{k}{2\pi}\int du\tilde{d}\langle A_{u},\delta\tilde{A}\rangle = \frac{k}{2\pi}\int du d\phi dr Tr[\partial_{r}(A_{u}\delta A_{\phi})-\partial_{\phi}(A_{u}\delta A_{r})]
\end{equation}
Apart from the boundary term, $\delta I_{H}[A]=0 \Longrightarrow \mathcal{F}_{r\phi}=0,\hspace{2pt} \mathcal{F}_{u r}=0, \hspace{2pt} \mathcal{F}_{\phi u}=0$ which means
\begin{equation}
\nonumber
F=dA+A^{2}=0
\end{equation}

\section{Details of the Computations for Dual WZW theory of ${\cal{N}}=(1,1)$ Supergravity}\label{AppC}
In this appendix, we shall briefly present an independent computation for the ${\cal{N}}=(1,1)$ case.  This is a simpler version of ${\cal{N}}=(2,0)$ case as we do not have any internal symmetry generators $T,Z$. But in calculations, the exact behaviour of fields (their overall signs) differ from the ${\cal{N}}=(2,0)$ case and also the basis of fermionic generators are different. Thus although the final result is mere a truncation of the  ${\cal{N}}=(2,0)$ one. The action is given in \eqref{action1}. We begin with eoms as, 
Equations of motion:\\
\begin{eqnarray}
de+[\hat{\omega},e]&=&\frac{1}{4}\sum_{i=1}^{2}(\overline{\psi}^{i}\psi^{i}-\half\overline{\psi}^{i}\psi^{i}\mathbb{I}), \hspace{21pt} d\hat{\omega}+\hat{\omega}^{2}=0\\
D\psi^{\alpha i}&=&-\half\gamma e^{a}(\Gamma_{a})^{\alpha}_{\beta}\psi^{\beta i}, \hspace{21pt} (i=1,2)
\end{eqnarray}
Solutions to equations of motion:\\
\begin{equation}
\hat{\omega}=\Lambda^{-1}d\Lambda, \hspace{21pt} \Lambda\in SL(2,\mathbb{R})
\end{equation}
\begin{equation}
\psi^{i}=\Lambda^{-1}d\eta^{i} \hspace{21pt} (i=1,2)
\end{equation}
\begin{equation}
e=\Lambda^{-1}(\frac{1}{4}\sum_{i=1}^{2}\eta^{i}d\overline{\eta}^{i}+\frac{1}{8}\sum_{i=1}^{2}d\overline{\eta}^{i}\eta^{i}\mathbb{I}+db)\Lambda
\end{equation}

Asymptotic form of the r-independent part of the gauge field in radial gauge:
\begin{equation}
A=(\frac{\mathcal{M}}{2}du + \frac{\mathcal{N}}{2}d\phi)P_{0} + du P_{1} + \frac{\mathcal{M}}{2}d\phi J_{0} + d\phi J_{1} + \sum_{i=1}^{2}\frac{\psi^{i}}{2^{1/4}}\mathcal{Q}^{i}_{+}
\end{equation}
Functional form of the solutions in radial gauge:
\begin{eqnarray}
\Lambda&=&\lambda(u,\phi)\xi(u,r)\\
\eta^{i\alpha}&=&\nu^{i\alpha}(u,\phi)+\lambda(u,\phi)\rho^{i\alpha}(u,r)\\
b&=&\alpha(u,\phi)-\frac{1}{4}\sum_{i=1}^{2}\nu^{i}(u,\phi)\overline{\rho}^{i}(u,r)\lambda^{-1}(u,\phi)-\frac{1}{8}\sum_{i=1}^{2}\overline{\rho}^{i}(u,r)\lambda^{-1}(u,\phi)\nu^{i}(u,\phi)\mathbb{I}\\
\nonumber
&&+\lambda(u,\phi)\beta(u,r)\lambda^{-1}(u,\phi)
\end{eqnarray}
Constraints on the asymptotic gauge field components:
\begin{equation}
\omega^{a}_{\phi}=e^{a}_{u}, \quad \omega^{a}_{u}=0, \quad \psi^{i +}_{u}=0=\psi^{i -}_{u}
\end{equation}
Surface term at the boundary:
\begin{equation}
-\frac{k}{2\pi}\int du \tilde{d}\langle A_{u},\delta\tilde{A}\rangle = -\frac{k}{4\pi}\int du d\phi\hspace{2pt} \omega^{a}_{\phi}\omega_{a \phi}|^{r=r_{0}}
\end{equation}
Action in terms of gauge field components:
\begin{eqnarray}
I&=&\frac{k}{4\pi}[\int du d\phi(e^{a}_{\phi}\omega_{a u}+\omega^{a}_{\phi}e_{a u}+\mu \omega^{a}_{\phi}\omega_{a\phi}-\sum_{i=1}^{2}\overline{\psi}^{i}_{\alpha u}\psi^{\alpha i}_{\phi})|^{r=r_{0}}\\
\nonumber
&&+\frac{1}{6}\int (3\epsilon^{abc}e_{a}\omega_{b}\omega_{c}+\mu \epsilon^{abc}\omega_{a}\omega_{b}\omega_{c}+\sum_{i=1}^{2}\frac{3}{2}\omega^{a}(C\Gamma_{a})_{\alpha\beta}\psi^{\alpha i}\psi^{\beta i})]
\end{eqnarray}
Action on the solutions of equations of motion:
\begin{equation}
I=\frac{k}{2\pi}(\int du d\phi \hspace{2pt} Tr[2\dot{\Lambda}\Lambda^{-1}(-\frac{1}{4}\sum_{i=1}^{2}\eta^{i}\overline{\eta}^{i '}+b')-(\Lambda'\Lambda^{-1})^{2}+\mu\Lambda'\Lambda^{-1}\dot{\Lambda}\Lambda^{-1}-\half\sum_{i=1}^{2}\eta^{i '}\dot{\overline{\eta}}^{i}]^{r=r_{0}}+\frac{\mu}{3}\int Tr(d\Lambda\Lambda^{-1})^{3})
\end{equation}
Action after using gauge decomposed forms of the solutions:
\begin{eqnarray}
I[\lambda,\alpha,\nu^{1},\nu^{2}]&=\frac{k}{2\pi}(\int du d\phi[2\dot{\lambda}\lambda^{-1}\alpha'+\half\sum_{i=1}^{2}\dot{\lambda}\lambda^{-1}\nu^{i}\overline{\nu}^{i '}-(\lambda' \lambda^{-1})^{2}+\mu\lambda' \lambda^{-1}\dot{\lambda}\lambda^{-1}-\half\sum_{i=1}^{2}\dot{\overline{\nu}}^{i}\nu^{i '}] \nonumber\\
&+\frac{\mu}{3}\int Tr(d\Lambda\Lambda^{-1})^{3})
\end{eqnarray}
Equations of motion:
\begin{eqnarray}
(\dot{\lambda}\lambda^{-1})'&=&0\\
\dot{\overline{\nu}}^{i '}+\overline{\nu}^{i '}\dot{\lambda}\lambda^{-1}&=&0\\
\dot{\alpha}'+\alpha'\dot{\lambda}\lambda^{-1}-\dot{\lambda}\lambda^{-1}\alpha'+\frac{1}{4}\sum_{i=1}^{2}\dot{\nu}^{i}\overline{\nu}^{i '}+\frac{1}{4}\sum_{i=1}^{2}\dot{\lambda}\lambda^{-1}\nu^{i}\overline{\nu}^{i '}-\partial_{\phi}(\lambda'\lambda^{-1})&=&0
\end{eqnarray}
Generic solutions of the equations of motion:
\begin{eqnarray}
\lambda&=&\tau(u)\kappa(\phi)\\
\nu^{i}&=&\tau(\zeta_{1}^{i}(u)+\zeta_{2}^{i}(\phi))\\
\alpha&=&\tau(a(\phi)+\delta(u)+u\kappa'\kappa^{-1}-\frac{1}{4}\sum_{i=1}^{2}\zeta_{1}^{i}\overline{\zeta}_{2}^{i})\tau^{-1}
\end{eqnarray}
Symmetries of the solutions:
\begin{eqnarray}
\alpha&\longrightarrow&\alpha+\lambda\Sigma(\phi)\lambda^{-1}\\
\lambda&\longrightarrow&\lambda\Theta^{-1}(\phi); \quad \alpha\longrightarrow\alpha-u\lambda\Theta^{-1}\Theta'\lambda^{-1}\\
\nu^{i}&\longrightarrow&\nu^{i}+\lambda\Upsilon^{i}(\phi) \quad \text{(i=1 or 2)}; \quad \alpha\longrightarrow\alpha-\frac{1}{4}\nu^{i}\overline{\Upsilon}^{i}\lambda^{-1}
\end{eqnarray}
Infinitesimal version of the symmetries:
\begin{eqnarray}
\delta_{\sigma}\alpha&=&\lambda\sigma(\phi)\lambda^{-1}\\
\delta_{\theta}\lambda&=&-\lambda\theta (\delta_{\theta}\lambda^{-1}=\theta\lambda^{-1}); \quad \delta_{\theta}\alpha=-u\lambda\theta'\lambda^{-1}\\
\delta_{\gamma}\nu^{1}&=&\lambda\gamma^{1}; \quad
\delta_{\gamma}\alpha=-\frac{1}{4}\nu^{1}\overline{\gamma}^{1}\lambda^{-1}\\
\delta_{\gamma}\nu^{2}&=&\lambda\gamma^{2}; \quad
\delta_{\gamma}\alpha=-\frac{1}{4}\nu^{2}\overline{\gamma}^{2}\lambda^{-1}
\end{eqnarray}
Currents corresponding to the above symmetries:
\begin{eqnarray}
J^{\mu}_{\sigma}&=&\delta^{\mu}_{0}(\frac{k}{\pi})Tr[\sigma\lambda^{-1}\lambda']=2\delta^{\mu}_{0}\sigma^{a}P_{a}\\
J^{\mu}_{\theta}&=&-\frac{k}{\pi}\delta^{\mu}_{0}Tr[\theta (\lambda^{-1}\alpha'\lambda-u(\lambda^{-1}\lambda')'+\frac{1}{4}\sum_{i=1}^{2}\lambda^{-1}\nu^{i}\overline{\nu}^{i'}\lambda)]=2\delta^{\mu}_{0}\theta^{a}J_{a}\\
J^{\mu}_{\gamma^{i}}&=&\frac{k}{2\pi}\delta^{\mu}_{0}Tr[\gamma^{i}\overline{\nu}^{i'}\lambda]=\delta^{\mu}_{0}Q^{i}_{\alpha}\gamma^{i\alpha} \quad (i=1,2, \quad \text{$i$ not summed over})
\end{eqnarray}
where $\sigma$ and $\theta$ being $SL(2,\mathbb{R})$ matrices are expanded in the basis of $\Gamma$ matrices.\\

Dirac brackets:
\begin{eqnarray}
\{P_{a}(\phi),P_{a}(\phi')\}&=&0\\
\{P_{a}(\phi),J_{b}(\phi')\}&=&\epsilon_{abc}P^{c}(\phi)\delta(\phi-\phi')-\frac{k}{2\pi}\eta_{ab}\delta'(\phi-\phi')\\
\{P_{a}(\phi),Q^{i}_{\alpha}(\phi')\}&=&0 \quad (i=1,2)\\
\{J_{a}(\phi),J_{b}(\phi')\}&=&\epsilon_{abc}J^{c}\delta(\phi-\phi')+\frac{\mu k}{2\pi}\eta_{ab}\delta'(\phi-\phi') \\
\{J_{a}(\phi),Q_{\alpha}^{i}(\phi')\}&=&\half(Q^{i}\Gamma_{a})_{\alpha}(\phi)\delta(\phi-\phi')\quad (i=1,2)\\
\{Q_{\alpha}^{i}(\phi),Q_{\beta}^{j}(\phi')\}&=&\delta^{ij}[-\half(C\Gamma^{a})_{\alpha\beta}P_{a}(\phi)\delta(\phi-\phi')-\frac{k}{2\pi}C_{\alpha\beta}\delta'(\phi-\phi')]
\end{eqnarray}
Bilinears for Sugawara construction:
\begin{equation}
\mathcal{H}=\frac{\pi}{k}P^{a}P_{a}, \quad \mathcal{P}=-\frac{2\pi}{k}J^{a}P_{a}+\mu\mathcal{H}+\frac{\pi}{k}\sum_{i=1}^{2}Q^{i}_{\alpha}C^{\alpha\beta}Q^{i}_{\beta}, \quad \mathcal{G}^{i}=2^{3/4}\frac{\pi}{k}(P_{2}Q^{i}_{+}+\sqrt{2}P_{0}Q^{i}_{-}) \quad (i=1,2)
\end{equation}
Remaining constraints on the gauge field:
\begin{equation}
\omega^{1}_{\phi}=1, \quad e^{1}_{\phi}=e^{2}_{\phi}=0, \quad \psi^{1-}_{\phi}=\psi^{2-}_{\phi}=0 
\end{equation}
Constraints on the fields:
\begin{equation}
\nonumber [\lambda^{-1}d\lambda]^{1}=1, \quad [\lambda^{-1}(\frac{1}{4}\sum_{i=1}^{2}\nu^{i}\overline{\nu}^{i'}+\frac{1}{8}\sum_{i=1}^{2}\overline{\nu}^{i'}\nu^{i}\mathbb{I}+\alpha')\lambda]^{1}=0, \quad [\lambda^{-1}\nu^{i'}]^{-}=0
\end{equation}
In terms of components of currents,
\begin{equation}
P_{0}(\phi)=\frac{k}{2\pi}, \quad J_{0}(\phi)=-\frac{\mu k}{2\pi}, \quad Q^{1}_{+}=Q^{2}_{+}=0
\end{equation}
Shifted bilinears:
\begin{equation}
\tilde{\mathcal{H}}=\mathcal{H}+\partial_{\phi}P_{2},\quad \tilde{\mathcal{P}}=\mathcal{P}-\partial_{\phi}J_{2}, \quad \tilde{\mathcal{G}}^{i}=\mathcal{G}^{i}+2^{3/4}\partial_{\phi}Q^{i}_{+}
\end{equation}
Poisson brackets:
\begin{eqnarray}
\{\tilde{\mathcal{H}}(\phi),\tilde{\mathcal{H}}(\phi')\}&=&0\\
\{\tilde{\mathcal{H}}(\phi),\tilde{\mathcal{P}}(\phi')\}&=&(\tilde{\mathcal{H}}(\phi)+\tilde{\mathcal{H}}(\phi'))\delta'(\phi-\phi')-\frac{k}{2\pi}\partial^{3}\delta(\phi-\phi')\\
\{\tilde{\mathcal{H}}(\phi),\tilde{\mathcal{G}}^{i}(\phi')\}&=&0\\
\{\tilde{\mathcal{P}}(\phi),\tilde{\mathcal{P}}(\phi')\}&=&(\tilde{\mathcal{P}}(\phi)+\tilde{\mathcal{P}}(\phi'))\delta'(\phi-\phi')-\frac{\mu k}{2\pi}\partial^{3}\delta(\phi-\phi')\\
\{\tilde{\mathcal{P}}(\phi),\tilde{\mathcal{G}}^{i}(\phi')\}&=&(\tilde{\mathcal{G}}^{i}(\phi)+\half\tilde{\mathcal{G}}^{i}(\phi'))\delta'(\phi-\phi')\\
\{\tilde{\mathcal{G}}^{i}(\phi),\tilde{\mathcal{G}}^{i}(\phi')\}&=&\delta^{ij}(\tilde{\mathcal{H}}(\phi)\delta(\phi-\phi')-\frac{k}{2\pi}\partial^{2}\delta(\phi-\phi'))
\end{eqnarray}

\section{Currents corresponding to Global symmetries of WZW theory}\label{AppD}

Here we present a procedure to get the $\phi$ dependent symmetries of the solutions with one example. Let us look at the solution of $\lambda$: $$\lambda= \tau(u)\kappa(\phi).$$ Multiplying the solution by an arbitrary $\phi$ dependent $SL(2,C)$ field $\theta^{-1}$ from right is still a symmetry of the solution. In this new solution $\kappa$ is modified as $ \kappa \theta$. Since $\kappa$ appears in the solution of $F$, that solution also needs to be transformed accordingly. The $\kappa$ dependent term in $F$ is $ \sim - u \tau \kappa' \kappa^{-1} \tau^{-1}$. This for $\kappa \rightarrow \kappa \theta$ this piece transforms as $ - u \tau (\kappa \theta^{-1})' (\kappa \theta^{-1})^{-1} \tau^{-1} = - u \tau \kappa' \kappa^{-1} \tau^{-1} + u \lambda ( \theta^{-1}\theta') \lambda^{-1}$. Thus the field $F$ changes as $F \rightarrow F + u \lambda (\theta^{-1}\theta') \lambda^{-1} $. A similar analysis for all possible symmetries of the solutions yields the transformations presented in the first equation of section \ref{sec5.1}. One can then easily derive the infinitesimal versions presented in \eqref{ist}.Below we present some details of the current computations. 

Noether current associated to a global symmetry generated by parameter $\epsilon$ is given as, 
\begin{equation}\label{D.1}
{\cal {J^{\mu}}}_{\epsilon}= \frac{\delta \cal{L}}{\delta(\partial_{\mu} \phi_i)}\delta_{\epsilon}\phi_i- K ^{\mu}_{\epsilon}, \qquad \partial_{\mu}K ^{\mu}_{\epsilon}= \delta_{\epsilon}\cal{L}.
\end{equation}
Another useful way to get the current is to use 
\begin{equation}\label{D.2}
\partial_{\mu}{\cal {J^{\mu}}}_{\epsilon}=\bigg( \frac{\delta \cal{L}}{\delta(\partial_{\mu} \phi_i)}-\frac{\delta \cal{L}}{\delta \phi_i}\bigg)\delta_{\epsilon}\phi_i.
\end{equation}

We shall write the current such that it only has non-zero component in the $u-$direction. For finding currents corresponding to $\mathcal{C}$ and $N$ transformations, \eqref{D.2} is useful and that directly gives us $J^{\mu}_C, J^{\mu}_N$ of \eqref{cc}. For the other four currents we need to use either of \eqref{D.1}, \eqref{D.2} and improvement terms $S^{[\mu \nu]}_{\epsilon}, T^{\mu}_{\epsilon}$. 

First we look at the fermionic currents. For these, we do not require improvement and \eqref{D.1} directly gives us currents in $u-$direction. In particular the for $J^{\mu}_{D_1}$, we get  $$K^{u}= (-\frac{k}{2\pi}) Tr[(\bar{d_2}' \lambda - i a' \bar{d_2} \lambda) D_1], \quad  K^{\phi}=\frac{k}{2\pi}Tr[\lambda D_1 \dot{\bar{d_2}}+ i \dot a \overline{\lambda D_1}d_2].$$ Finally for $J^{\mu}_{D_1}$ we need to take contribution from $\frac{\delta \cal{L}}{\delta(\partial_{\mu} \phi_i)}\delta_{\epsilon}\phi_i$ piece, that cancels the $K^{\phi}$ part and adds an equal contribution as $K^u$ in the final expression. Similarly we get $J^{\mu}_{D_2}$.

For the current due to $A$ transformation, direct evaluation with \eqref{D.1} gives $${\cal{J}}^{u}= \frac{k}{4\pi} [(\bar{\mu} a' + 2i c'+2i(\bar{d_2}' d_1 - \bar{d_1}' d_2 - ia' (\bar{d_2}d_1 + \bar{d_1} d_2)))A-\bar \mu a A'+ 8 u a' A'] , \\ {\cal {J^{\phi}}}=\frac{k}{4\pi}(\bar \mu \dot a A - 8 a' A) .$$ Here the fermion terms are traced among themselves. To get the final current in $u-$direction, we need to add $$S^{[\mu \nu]}_A= -\frac{k}{4\pi}\varepsilon^{\mu \nu}(8 u a'A-\bar \mu a A), \quad \varepsilon^{u \phi}=1, \quad T^{\mu}_A=\delta ^{\mu}_{\phi} u \frac{k}{4\pi}\dot{a'}.$$ 
taking these improvement terms into account we finally get $J^{\mu}_{A}$ as given in \eqref{cc}. 

Finally for $\Theta$ transformation, using  \eqref{D.2} we get $${\cal {J}}^{u}=\frac{k}{2\pi} Tr[\{\lambda^{-1}\hat\alpha \lambda+ \mu \lambda^{-1}\lambda' \}\theta-2u (\lambda^{-1}\lambda')\theta'], \quad {\cal {J^{\phi}}}=\frac{k}{2\pi}Tr[ \theta \lambda^{-1} \lambda']$$. Adding the required improvements terms are $$S^{[\mu \nu]}_{\Theta}=  \frac{k}{2\pi}\varepsilon^{\mu \nu}Tr[u \Theta \lambda^{-1} \lambda'], \quad T^{\mu}_{\Theta}= \delta ^{\mu}_{\phi} u \frac{k}{2\pi} Tr[\Theta (\lambda^{-1} \lambda')^{\bf{\cdot}}], $$ we finally get the expression for $J^{\mu}_{\Theta}$ as given in \eqref{cc}. 

\section{Some important Dirac Brackets}\label{AppE}

In this appendix, we provide the nontrivial Dirac brackets between various currents and current bilinears that are required for the results presented in the draft.

The non-trivial Dirac brackets of Sugawara modes with currents are given by:
\begin{align}
\left\lbrace H(\phi), Q^J_a(\phi')  \right\rbrace &= - Q^P_a(\phi) \delta'(\phi-\phi')\\
\left\lbrace H(\phi), Q^A(\phi')  \right\rbrace &= 4  Q^C(\phi) \delta'(\phi-\phi')\\
\left\lbrace P(\phi), Q^J_a(\phi')  \right\rbrace &=  Q^J_a(\phi) \delta'(\phi-\phi')\\
\left\lbrace P(\phi), Q^P_a(\phi')  \right\rbrace &=  Q^P_a(\phi) \delta'(\phi-\phi')\\
\left\lbrace P(\phi), Q^{G_1}_{\gamma}(\phi')  \right\rbrace &=  Q^{G_1}_{\gamma}(\phi) \delta'(\phi-\phi')\\
\left\lbrace P(\phi), Q^{G_2}_{\gamma}(\phi')  \right\rbrace &= Q^{G_2}_{\gamma}(\phi) \delta'(\phi-\phi')\\
\left\lbrace P(\phi), Q^A(\phi')  \right\rbrace &=  Q^A(\phi) \delta'(\phi-\phi')\\
\left\lbrace P(\phi), Q^C(\phi')  \right\rbrace &=  Q^C(\phi) \delta'(\phi-\phi')\\
\left\lbrace \mathcal{G}^1(\phi), Q^{G_2}_{+}(\phi')  \right\rbrace &= \sqrt{2} \frac{\pi}{k} Q^P_0 \frac{k}{\pi}  \delta'(\phi-\phi')\\
\left\lbrace \mathcal{G}^1(\phi), Q^{G_2}_{-}(\phi')  \right\rbrace &=  H(\phi) \delta(\phi-\phi') - Q^P_2(\phi) \delta'(\phi-\phi') -2i Q^C(\phi) \delta'(\phi-\phi')\\
\lbrace  \mathcal{G}_1(\phi), Q^A(\phi)  \rbrace &= -i \mathcal{G}_1(\phi) \delta (\phi-\phi')\\
\lbrace  \mathcal{G}_2(\phi), Q^A(\phi)  \rbrace &= i \mathcal{G}_2(\phi) \delta (\phi-\phi')
\end{align}

With above equations we can try to calculate the PBs between different modes of stress tensor. For example:
\begin{align*}
\left\lbrace H(\phi), P(\phi')\right\rbrace &= \left\lbrace H(\phi), (- 2 \frac{\pi}{k}) Q^J_a Q^P_a\right\rbrace\\
&= (- 2 \frac{\pi}{k}) [\left\lbrace H(\phi), Q^J_a (\phi') \right\rbrace Q^P_a (\phi') + Q^J_a (\phi') \left\lbrace H(\phi), Q^P_a (\phi')\right\rbrace Q^P_a (\phi')]\\
&=  2 \frac{\pi}{k} Q^P_a(\phi) Q^P_a (\phi') \partial_{\phi}\delta(\phi-\phi') \\
&= \frac{\pi}{k} Q^P_a(\phi) \partial_{\phi}[Q^P_a (\phi') \delta(\phi-\phi')] - \frac{\pi}{k} \partial_{\phi'}[Q^P_a(\phi)\delta(\phi-\phi')]  Q^P_a (\phi') \\
&= \frac{\pi}{k} Q^P_a(\phi) \partial_{\phi}[Q^P_a (\phi) \delta(\phi-\phi')] - \frac{\pi}{k} \partial_{\phi'}[Q^P_a(\phi')\delta(\phi-\phi')]  Q^P_a (\phi') \\
&= \frac{\pi}{k} Q^P_a(\phi) Q^P_a (\phi) \partial_{\phi} \delta(\phi-\phi') - \frac{\pi}{k} \partial_{\phi'}\delta(\phi-\phi') Q^P_a(\phi') Q^P_a (\phi') \\
&= (H(\phi) + H(\phi')) \partial_{\phi} \delta(\phi-\phi')
\end{align*}

The Dirac Brackets of the above modes among themselves are given by:
\begin{align}
\lbrace H(\phi), H(\phi') \rbrace &= 0\\
\lbrace H(\phi), P(\phi') \rbrace &= (H(\phi) + H(\phi')) \partial_{\phi} \delta(\phi - \phi')\\
\lbrace P(\phi), P(\phi') \rbrace &= (P(\phi) + P(\phi')) \partial_{\phi} \delta(\phi - \phi')\\
\lbrace P(\phi), \mathcal{G}^I(\phi') \rbrace &= (\mathcal{G}^I(\phi) + \mathcal{G}^I(\phi'))\partial_{\phi} \delta(\phi - \phi')\\
\lbrace H(\phi), Q^A(\phi') \rbrace &= 4 Q^C(\phi) \partial_{\phi} \delta(\phi - \phi')\\
\left\lbrace  Q_C(\phi), Q_A(\phi')    \right\rbrace &=  \frac{k}{2\pi} \partial_{\phi} \delta(\phi-\phi')\\
\lbrace P(\phi), Q^C(\phi') \rbrace &= Q^C(\phi)\partial_{\phi} \delta(\phi - \phi')\\
\lbrace P(\phi), Q^A(\phi') \rbrace &= Q^A(\phi) \partial_{\phi}\delta(\phi-\phi')\\
\lbrace  \mathcal{G}_1(\phi), Q^A(\phi)  \rbrace &= -i \mathcal{G}_1(\phi) \delta (\phi-\phi')\\
\lbrace  \mathcal{G}_2(\phi), Q^A(\phi)  \rbrace &= i \mathcal{G}_2(\phi) \delta (\phi-\phi')\\
\end{align}

The modes of Stress tensor as defined by above Sugawara construction do not commute with the First class constraints. In fact,
\begin{align*}
\lbrace H(\phi), Q^P_0(\phi')  \rbrace &=0\\
\lbrace H(\phi), Q^J_0(\phi')  \rbrace &= -Q^P_0(\phi) \delta'(\phi-\phi') = -\sqrt{2} \frac{k}{2\pi} \delta'(\phi-\phi')\\
\lbrace H(\phi), Q^{G_1}_+(\phi')  \rbrace &= \lbrace H(\phi), Q^{G_2}_+(\phi')  \rbrace = 0\\
\lbrace P(\phi), Q^P_0(\phi')  \rbrace &= Q^P_0(\phi) \delta'(\phi-\phi') = \sqrt{2} \frac{k}{2\pi} \delta'(\phi-\phi')\\
\lbrace P(\phi), Q^J_0(\phi')  \rbrace &= Q^J_0(\phi) \delta'(\phi-\phi')= \sqrt{2} \frac{\mu k}{4\pi} \delta'(\phi-\phi')\\
\lbrace P(\phi), Q^{G_{1,2}}_+(\phi')  \rbrace &= Q^{G_{1,2}}_+(\phi) \delta'(\phi-\phi') =0\\
\lbrace  \mathcal{G}^{1,2}_{\alpha}(\phi) , Q^P_0(\phi') \rbrace &= 0\\
\lbrace  \mathcal{G}^{1,2}_{\alpha}(\phi) , Q^J_0(\phi') \rbrace &= Q^P_0(\phi) Q^{1,2}_+(\phi) \delta(\phi-\phi')\\
\left\lbrace \mathcal{G}^1(\phi), Q^{G_2}_{+}(\phi')  \right\rbrace &= \sqrt{2} \frac{\pi}{k} Q^P_0 \frac{k}{\pi}  \delta'(\phi-\phi') =\frac{k}{\pi} \delta'(\phi-\phi') \\
\lbrace Q^A(\phi), Q^{G_1}_+(\phi')  \rbrace &= i Q^{G_1}_+(\phi) \delta(\phi-\phi')=0\\
\lbrace Q^A(\phi), Q^{G_2}_+(\phi')  \rbrace &= - i Q^{G_2}_+(\phi) \delta(\phi-\phi') =0
\end{align*}

\section{An example of Super BMS$_3$ current commutation}\label{AppF}
In this appendix we shall present how the shifted fermionic currents $ \hat{\mathcal{G}}^1 (\phi), \hat{\mathcal{G}}^2 (\phi')   $ closes to right SuperBMS$_3$ structure under anti-commutation. 
With these shifts the Dirac bracket becomes:
\begin{align*}
\lbrace \hat{\mathcal{G}}^1 (\phi), \hat{\mathcal{G}}^2 (\phi') \rbrace =& \lbrace \mathcal{G}^1 (\phi)+ \partial_{\phi} Q^{G_1}_+ (\phi), \mathcal{G}^2 (\phi') + \partial_{\phi'} Q^{G_2}_+(\phi') \rbrace \\
=& \lbrace \mathcal{G}^1 (\phi), \frac{\pi}{k} (Q^P_2 Q^{G_2}_+ + \sqrt{2} Q^P_0 Q^{G_2}_- + 2i Q^C Q^{G_2}_+) \rbrace + \partial_{\phi} \lbrace Q^{G_1}_+ (\phi), \mathcal{G}^2 (\phi') \rbrace\\
&+ \partial_{\phi'} \lbrace \mathcal{G}^1 (\phi) ,Q^{G_2}_+(\phi')\rbrace + 
\partial_{\phi'} \partial_{\phi} \lbrace  Q^{G_1}_+(\phi),Q^{G_2}_+(\phi')  \rbrace
\end{align*}
Now we can look at the RHS term by term. The first DB gives:
\begin{align*}
&\lbrace \mathcal{G}^1 (\phi), \frac{\pi}{k} (Q^P_2 Q^{G_2}_+ + \sqrt{2} Q^P_0 Q^{G_2}_- + 2i Q^C Q^{G_2}_+) (\phi')\rbrace\\
=& \frac{\pi}{k} Q^P_2 (\phi') \lbrace \mathcal{G}^1 (\phi), Q^{G_2}_+  (\phi') \rbrace + \frac{\pi}{k} \sqrt{2} Q^P_0  (\phi')\lbrace \mathcal{G}^1 (\phi), Q^{G_2}_-  (\phi')\rbrace + 2i \frac{\pi}{k} \lbrace \mathcal{G}^1 (\phi),Q^{G_2}_+  (\phi')\rbrace
\end{align*}
Using the fact that we are on the constrained surface defined by \eqref{2cc}, we see that the first term above is 0 since $Q^P_2(\phi) =0$ on the surface. The last two terms combine to give:
\begin{align*}
&\lbrace \mathcal{G}^1 (\phi), \frac{\pi}{k} (Q^P_2 Q^{G_2}_+ + \sqrt{2} Q^P_0 Q^{G_2}_- + 2i Q^C Q^{G_2}_+) (\phi')\rbrace = \mathcal{H}(\phi)\delta(\phi-\phi') - 2i (Q^C(\phi)+Q^C(\phi'))\delta'(\phi-\phi')
\end{align*}
Where we also needed to use $Q^P_2(\phi) \partial_{\phi} \delta(\phi-\phi') = -\partial_{\phi} Q^P_2(\phi)\delta(\phi-\phi') $ on the constrained surface.

Similarly the last rest of the terms of the first DB combine to give: 
\begin{align*}
\partial_{\phi} \lbrace Q^{G_1}_+ (\phi), \mathcal{G}^2 (\phi') \rbrace
+ \partial_{\phi'} \lbrace \mathcal{G}^1 (\phi) ,Q^{G_2}_+(\phi')\rbrace + 
\partial_{\phi'} \partial_{\phi} \lbrace  Q^{G_1}_+(\phi),Q^{G_2}_+(\phi')  \rbrace = -\frac{k}{\pi} \partial^2_{\phi} \delta(\phi - \phi')
\end{align*}
Thus we finally get:
\begin{align*}
\lbrace \hat {\mathcal{G}}^1(\phi),\hat {\mathcal{G}}^2(\phi')  \rbrace_{DB} &=  {\cal{H}}(\phi)\delta(\phi-\phi')-\frac{k}{\pi}\partial^2_{\phi}\delta(\phi-\phi') -2i(Q^C(\phi)+Q^C(\phi'))\delta'(\phi-\phi')
\end{align*}

Similar computations can be performed with other shifted currents to get the final Dirac brackets.

\bibliography{bms3}
\bibliographystyle{jhep}

\end{document}

%% file: sugracom.tex
\newcommand {\cN}{{\cal N}}

\def\G{\Gamma}

\newcommand{\be}{\begin{equation}}
\newcommand{\ee}{\end{equation}}
\newcommand{\bea}{\begin{eqnarray}}
\newcommand{\eea}{\end{eqnarray}}

\newcommand{\ba}{\begin{array}}
\newcommand{\ea}{\end{array}}

\def\double #1{#1{\hbox{\kern-2pt $#1$}}}

\newcommand{\bsubeq}{\begin{subequations}}
\newcommand{\esubeq}{\end{subequations}}

